\documentclass[11pt]{article}
\usepackage[utf8]{inputenc}
\usepackage[T1]{fontenc}
\usepackage[english]{babel}

\usepackage{amsmath}
\usepackage{amssymb}
\usepackage{amsthm}
\usepackage{verbatim}
\usepackage{enumerate}
\usepackage{appendix}
\usepackage{geometry}
\usepackage{eurosym}
\usepackage{hyperref}
\hypersetup{
    colorlinks,
    citecolor=blue,
    filecolor=blue,
    linkcolor=blue,
    urlcolor=blue
}
\usepackage{todonotes}
\usepackage{mathrsfs}

 \usepackage{rotating}
\usepackage[symbol]{footmisc}

\def\be{\begin{eqnarray}}
\def\ee{\end{eqnarray}}

\def\b*{\begin{eqnarray*}}
\def\e*{\end{eqnarray*}}

\newtheorem{Theorem}{Theorem}[part]

\newtheorem{Proposition}{Proposition}[part]

\newtheorem{Lemma}{Lemma}[part]

\newtheorem{Remark}{Remark}[part]

\newcommand{\ba}{\begin{array}}
\newcommand{\ea}{\end{array}}
\newcommand{\ben}{\begin{equation*}}
\newcommand{\een}{\end{equation*}}
\newcommand{\bea}{\begin{eqnarray}}
 \newcommand{\eea}{\end{eqnarray}}
\newcommand{\bean}{\begin{eqnarray*}}
\newcommand{\eean}{\end{eqnarray*}}
\newcommand{\bel}{\begin{align}}
\newcommand{\eel}{\end{align}}
\newcommand{\beln}{\begin{align*}}
\newcommand{\eeln}{\end{align*}}
\newcommand{\bit}{\begin{itemize}}
\newcommand{\eit}{\end{itemize}}

\makeatletter \@addtoreset{equation}{section}

\@addtoreset{Definition}{section}

\@addtoreset{Theorem}{section}

\@addtoreset{Proposition}{section}

\@addtoreset{Property}{section}

\@addtoreset{Assumption}{section}

\@addtoreset{Corollary}{section}

\@addtoreset{Lemma}{section}

\@addtoreset{Remark}{section}

\@addtoreset{Example}{section}

\@addtoreset{Condition}{section}

\def \E{\mathbb{E}}

\def \H{\mathbb{H}}
\def \L{\mathbb{L}}
\def \M{\mathbb{M}}
\def \N{\mathbb{N}}
\def \P{\mathbb{P}}
\def \Q{\mathbb{Q}}
\def \R{\mathbb{R}}

\def \Z{\mathbb{Z}}

\def \G{\mathbb{G}}

\def\gik{\gamma^{i,j}}
\def\g1{\gamma^{i,1}}

\def\={\;=\;}
\def\.{\;.}
\def\a{\alpha}
\def\b{\beta}

\def\1{{\bf 1}}

 \def\normeL2#1{\left\|{#1}\right\|_{L^2}}

\newcommand{\alias}[2]{
\providecommand{#1}{}
\renewcommand{#1}{#2}
}
\alias{\E}{\mathbb{E}}

\alias{\P}{\mathbb{P}}
\alias{\N}{\mathcal{N}}
\alias{\L}{\mathcal{L}^2}
\alias{\Z}{\mathbb{Z}}
\alias{\Q}{\mathbb{Q}}
\alias{\R}{\mathbb{R}}
\alias{\C}{\mathcal{C}}
\alias{\T}{\mathbb{T}}
\alias{\E}{\mathbb{E}}
\alias{\H}{\mathcal{H}}
\alias{\1}{\mathbf{1}}
\alias{\B}{\mathcal{B}}
\alias{\M}{\mathcal{M}}
\alias{\G}{\mathcal{G}}
\alias{\Y}{Y_{\bullet}}

\newcommand{\nc}{\newcommand}
\nc{\cA}{{Y}} \nc{\cB}{{\mathcal B}} \nc{\cC}{{\mathcal
C}} \nc{\cD}{{\mathcal D}} \nc{\bbD}{\mathbb{D}}
\nc{\cG}{{\mathcal G}} \nc{\cF}{{\mathcal F}} \nc{\cS}{{\mathcal
S}} \nc{\cU}{{\mathcal U}} \nc{\cH}{{\mathcal H}}
\nc{\cK}{{\mathcal K}} \nc{\cM}{{\mathcal M}} \nc{\cO}{{\mathcal
O}} \nc{\cP}{{\mathcal P}} \nc{\bbE}{\mathbb{E}}
\nc{\bbEP}{\mathbb{E}_{\mathbb{P}}}\nc{\bbL}{\mathbb{L}}
\nc{\bbP}{\mathbb{P}} \nc{\bbQ}{\mathbb{Q}} \nc{\del}{\partial}
\nc{\Om}{\Omega} \nc{\om}{\omega} \nc{\bbR}{\mathbb{R}}
\nc{\bbC}{\mathbb{C}} \nc{\bfr}{\begin{flushright}}
\nc{\efr}{\end{flushright}} \nc{\dXt}{\Delta X_{t}}
\nc{\dXs}{\Delta X_{s}} \nc{\bs}{\blacksquare} \nc{\dX}{\Delta X}
\nc{\dY}{\Delta Y}
\nc{\dnkx}{\left(X(T^{n}_{k})-X(T^{n}_{k-1})\right)}
\nc{\esssup}{\mathrm{ess}\mbox{ }\mathrm{sup}}
\nc{\essinf}{\mathrm{ess}\mbox{ } \mathrm{inf}}
\nc{\dhats}{\widehat{\delta_s}}

\nc{\chf}{\mbox{$\mathbf1$}}
\nc{\ind}{\mathds{1}}

\begingroup\expandafter\expandafter\expandafter\endgroup
\expandafter\ifx\csname pdfsuppresswarningpagegroup\endcsname\relax
\else
\fi

\nc{\pinf}{P^{\infty} }
\nc{\aiopt}{\hat{\alpha}^i}
\nc{\biopt}{\hat{\beta}^i}

\nc{\mi}{{M_t^{a^i}}}
\nc{\mj}{{M_t^{a^j}}}
\nc{\gl}{{\lambda}}

\newcommand{\TL}{L^2(\Omega \times [0,T], (\mathcal{F}_t)_t, \P\otimes dt)}

\begin{document}
{\stepcounter{footnote}
\title{Optimal dynamic regulation of carbon emissions market \\ A variational approach}
\author{\stepcounter{footnote}René Aïd\thanks{University Paris-Dauphine, PSL Research University, Department of Economics, Place du Marechal de Lattre de Tassigny 75 775 Paris, France; rene.aid@dauphine.psl.eu. This project received the support from the Finance For Energy Markets Research Initiative (\url{https://www.fime-lab.org}) and the EcoRESS ANR under grant ANR-19-CE05-0042.}
 \quad Sara Biagini \thanks{ sbiagini@luiss.it, Department of Economics and Finance, LUISS University, viale Romania 32, 00197 Rome,  Italy. Financial support from LUISS Visiting Program is gratefully acknowledged.}}
\maketitle

\begin{abstract}
We consider the problem of  reducing the carbon emissions of a set of firms over a finite horizon. A regulator dynamically  allocates emission allowances to each firm. Firms face idiosyncratic as well as common economic shocks on emissions, and have linear quadratic abatement costs. Firms can trade allowances so to minimise total expected costs, from  abatement and trading  plus a quadratic terminal penalty. Using variational methods, we exhibit \emph{in closed-form} the market equilibrium in function of regulator's dynamic allocation. We then solve the  Stackelberg game between the regulator and the firms. Again, we obtain a   closed-form expression of the dynamic allocation policies that allow  a desired  expected emission reduction. Optimal policies are not unique but share common properties. Surprisingly, all optimal  policies induce a constant abatement effort and a \emph{constant price of allowances}. Dynamic allocations outperform static ones because of adjustment costs and uncertainty, in particular given the presence of common shocks. Our results are robust to some extensions, like risk aversion of firms or different penalty functions.

\vspace{5mm}




\noindent{\bf Acknowledgements:} We warmly thank Bruno Bouchard, Anna Creti, Paolo Guasoni,  Peter Tankov and  Nizar Touzi for discussions on the topic.
\end{abstract}

\section{Introduction}

Since its inception in 2005, the European Union Trading System has been a major innovative tool to manage carbon emissions and help EU member states reach their agreed reduction targets. Former examples of cap-and-trade mechanism to reduce pollution include the successful cap-and-trade market for sulfur dioxide and nitrogen oxides in the US (Title IV of Clean Air Act Amendments, 1990). However, the striking novelty of the EU carbon market is the dimension:  more than 30 billions euro of value gathering more than 15 thousands stationary installations across 30 countries. The seminal paper by Montgomery (1972) \cite{Montgomery72} on market for licences  has found here a spectacular illustration of the idea that market mechanisms can be efficiently developed to achieve pollution reduction. Nevertheless, after 15 years, it is clear that the EUTS is facing some issues. Figure~\ref{fig:carbonprice} provides the price of allowances from January, 2008 to June, 2020 and the cumulative difference between total verified emissions per year and total allowances. It shows that the market price of carbon is highly sensitive to the relation between  supply (allowances) and demand (emissions). The 2008 financial crisis led a large surplus that lasted until 2013 and led to a depressed market price, which reached  less than 5 \euro/tCO2. During this period, emissions were reduced, not because of firms abatement efforts, but  because the world was experiencing a major recession. This phenomenon led the EU to design the Market Stability Reserve mechanism (MSR), to reduce  market price volatility and over supply (EU Decision 2015/1814 of October 6${^\text{th}}$, 2015 on EU Directive 2003/87). In a nutshell, the MSR regulates   the potential market imbalances either by backloading allowances to the future or providing more allowances through auctions during the current phase and the next ones. This mechanism makes the carbon market regulation a dynamic process.

\begin{figure}[!ht]
\label{fig:carbonprice}
\center
\includegraphics[width=0.6\textwidth]{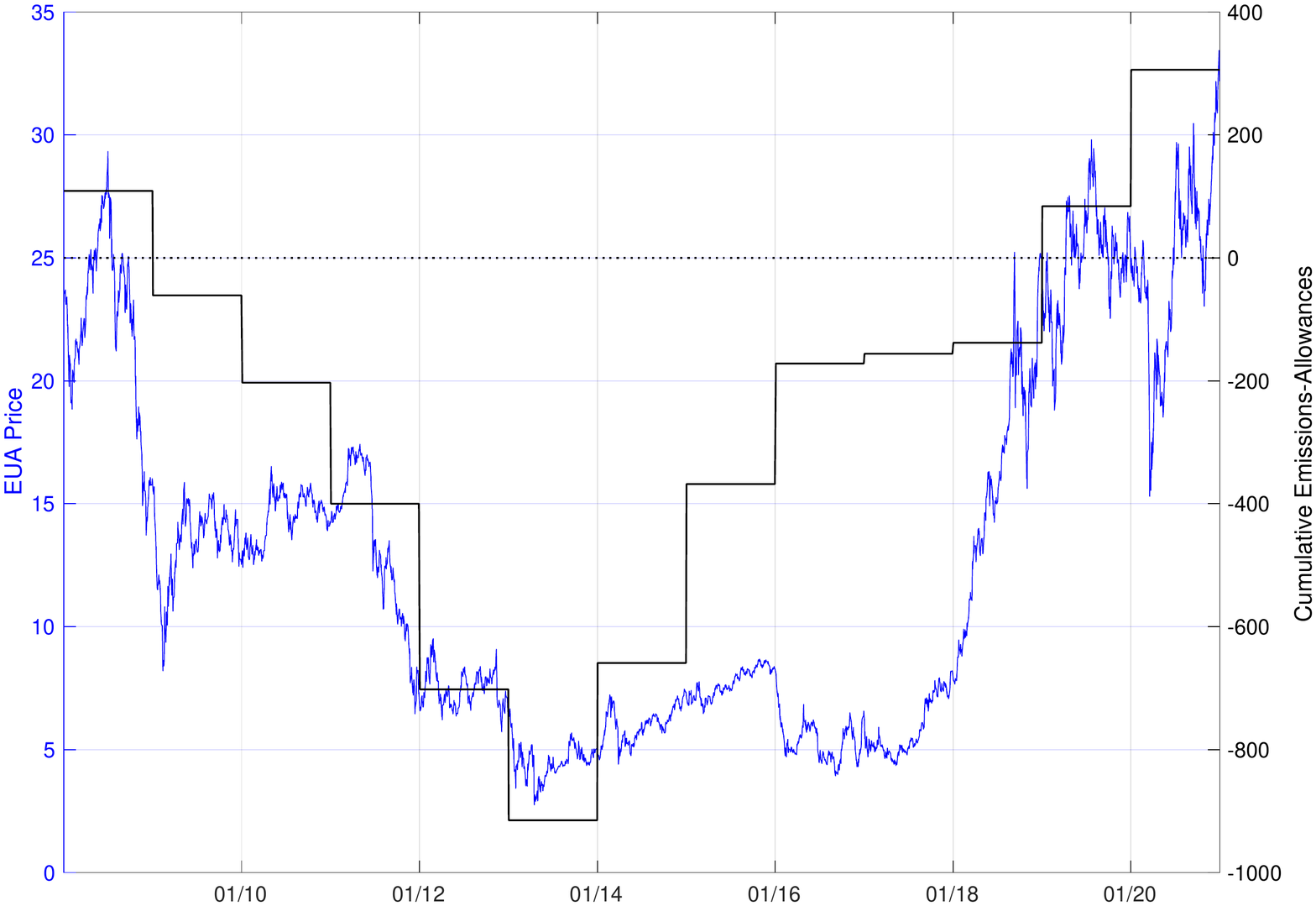}
\caption{EUA price in \euro/tCO2 (left axis)  compared to the difference between total verified emissions and total allocations in MtCO2-equiv (right axis) from January 2008 to June 2020; source: Eikon Reuters for EUA prices and European Environment Agency EU Emission Trading System data viewer.}.
\end{figure}

Thus, the MSR mechanism  implements the idea that contingent regulation should be preferred to fixed ones. This concept has been vastly supported in the literature, since  Weitzman's (1974) \cite{Weitzman74} seminal paper on regulation under uncertainty. From there, an intense research activity has focused on dynamic regulation in different settings. Hahn (1989) \cite{Hahn89} and more recently Hepburn (2006) \cite{Hepburn06} both provide  surveys on regulation through prices or quantities as well as a framing of the admissible tools to regulators. \\
\indent In our specific case of carbon  market, there exists an extensive literature which analyzes the carbon emissions reduction problem under its many different aspects. The optimality of banking permits from one period to the next has been studied in Rubin (1996) \cite{Rubin96}, Schennach (2000) \cite{Schennach00}, Chaton et al. (2015) \cite{Chaton15}, Lintunen and Kuusela  (2018) \cite{Lintunen18} and Kuusela and Lintunen (2020) \cite{Kuusela20}. Mechanism designs are proposed in Carmona et al. (2010) \cite{Carmona10} in a discrete time model to reduce electricity producers windfall profits. An important stream of the literature on diffusive pollution deals with imperfect competition and strategic interaction among polluters. Requate (1993) \cite{Requate93}, Von Der Fehr (1993) \cite{VonderFehr93} and recently Anand and Giraud-Carrier (2020) \cite{Anand20}  develop static models of imperfect competition which point out the firms capacity to increase their profit with emissions regulation. \\
\indent The focus in the present paper  is continuous time regulation of the carbon emissions with dynamic allowances allocation, see also Kollenberg and Taschini (2016, 2019), Grüll and Taschini (2011) \cite{Grull11}, Pizer (2002) \cite{Pizer02} and Pizer and Prest (2020) \cite{Pizer20}. However, we leave for future research the coupling of the dynamical aspects of emissions reductions with imperfect competition among polluters.
  We regard carbon reduction as a stochastic Stackelberg game. Firms are the Followers, and the regulator is the Leader. The model for firms is a continuous time stochastic model,  largely inspired by  the Kollenberg and Taschini (2016, 2019) \cite{Kollenberg16,Kollenberg19} carbon emissions market model. Firms  experience individual emissions growth rate, plus  idiosyncratic as well as common economic shocks.  Abatement costs are heterogeneous and  linear-quadratic. Although there are significant uncertainties on the marginal abatement costs of carbon reduction (see Gillingham and Stock (2018) \cite{Gillingham18} for an introduction on the topic), we make the assumption that at the time scale of carbon market emissions phase, less than ten years, these costs are known and constant. Firms can trade the carbon emissions allowances provided by the regulator, and emissions market imperfection is taken into account by an impact on the price of carbon. The key feature is that each firm is endowed with a bank account, as in the cited papers by Kollenberg and Taschini. The bank account position is the result of the allocation received, the traded permits minus the realised emissions. Further, firms face a terminal quadratic penalty on  their  bank position at the end of the regulated period - this is a strong incentive to achieve emission reduction. There are quite large uncertainties on the damage cost function induced by carbon emissions (see Hsiang and Kopp's \cite{Hsiang18}  and Auffhammer \cite{Auffhammer18} papers for a thorough introduction to  the uncertainties in carbon emissions damage function). Nevertheless, the convex terminal penalty can be interpreted as the expected future damage costs induced by non-compliant emissions. The objective of the authority is to achieve a given expected carbon emission reduction at least possible expected total cost from abatement, trading  and terminal penalty. Dynamic allocation processes available to the regulator are chosen in the  space of semimartingales, i.e. processes which are the sum of a diffusion part and a jump component.

Usually,  carbon emissions equilibrium price is tackled using forward-backward stochastic differential equations as   {illustrated in Carmona et al.} (2009, 2010, 2011, 2013) \cite{Carmona09,Carmona10,Carmona11,Carmona13}.   Further, stochastic dynamic Stackelberg games are typically solved using Bellman's principle and coupled HJB equations. On this, the reader is referred to Bensoussan et. al. (2014) \cite{Bensoussan14} for a survey on the problem and illustrations in the linear quadratic case. \\
\indent Instead, our approach is in the spirit of Duffie's utility maximisation via stochastic gradient methods (see Duffie (2001) \cite{Duffie01}).  Since  the objective functions in our problem are linear quadratic, they are Fréchet differentiable over the space of square integrable processes. This smoothness allows for stochastic variational techniques and leads to explicit and compact expressions   for the equilibrium price and optimal controls. In fact, using the variational approach  we provide closed form expressions for the best response of each firm. Without surprise, the optimal abatement effort equates the marginal abatement cost to the marginal penalty.  More surprisingly,  the abatement effort process is a martingale irrespectively of the allocation. We obtain in closed form a unique market equilibrium, reached at the trading rates that clear the market. The corresponding equilibrium price $\hat P$ is a martingale given by the conditional expectation of  the (average of) marginal penalties. Its dynamics are driven by (conditional) expectation of global  allocations and emission shocks. Of course, the higher the  allocations, the lower the market price, and conversely for shocks. Further, we show that the same methodology can be applied to different situations, like risk-averse firms or alternative terminal penalty function.

Our main findings consist in showing that optimal dynamic policies are not unique, though sharing common properties. All optimal dynamic policies induce a constant abatement effort. As a consequence of constant abatement efforts, the market equilibrium price is also constant. All dynamic allocations provide the same expected allocation. A simple example of optimal dynamic policy consists in, first, debiting the firms accounts by the level of the desired emission reduction and then to credit them at the business-as-usual emission rate, including  shocks. By acting in this way, the regulator {\em hedges} or {\em protects} firms against heavy adjustment costs. Far from being a drawback, the non-uniqueness of the optimal dynamic allocation  suggests that by the same tool  the regulator could reach other goals on top of  the reduction of expected emission at the minimum  social cost.  An example is adding  constraints on the financing of new technologies. We leave this analysis to future research.   \\
\indent Indeed, an   {accurate} comparison with a static allocation mechanism as implemented in the first phases of the EUTS shows that dynamic allocation mechanisms outperform static allocations only in the presence of  adjustment costs, business cycles and, in particular, of common shocks. If any of these three features is absent, there is no benefit from the implementation of a dynamic allocation mechanism. Using crude data of verified emission of the industrial sectors involved in the EUTS on the period 2008 to 2012, we calibrated our model.  Using the closed formulas then, we calculated the difference in cost between the optimal dynamic allocation described above and three alternative existing policies, namely the static allocation mechanism that prevailed during Phase~I and II, a MSR-like mechanism and the pure tax policy. We find significant difference in cost between static and dynamic allocations, approximately $20\%$ when flexibility is low. On the other hand, the pure tax policy induces costs of higher order of magnitude.

The paper is organised as follows. Section~\ref{sec:model} describes the stochastic underlying  model and formalises the regulator's problem. Section~\ref{sec:firm} deals with the firm individual optimisation problem for a fixed allocation process and a given market price process. Section~\ref{sec:market} provides the market equilibrium both with and without market impact. Section~\ref{sec:optimalregulation} solves the regulator's problem.  Comparisons with alternative policies, like Market Stability Reserve, pure tax and static allocation are gathered in Section~\ref{ssec:altpol}. There, numerical illustrations are also given. Finally, the Appendix contains technical results and computationally intensive parts of the proofs of the main Theorems.

\section{The model}
\label{sec:model}

 The regulation of carbon allowances is indeed quite complex. In particular, it occurs over several periods,  and allowances can be  banked  from one period to the other.   We abstract from these features and focus on a single period of $T$ years at the end of which compliance is assessed, as in Carmona et al. (2010, 2013) \cite{Carmona10,Carmona13}, Kollenberg and Taschini (2016, 2019) \cite{Kollenberg16,Kollenberg19}, Fell and Morgenstern (2010) \cite{Fell10}. \\
  \indent A regulator (the Leader) wishes to reduce the pollution produced by a set of $N$ firms (Followers) over a period  $[0,T]$. To this end, she allocates carbon permits to the firms. Given the received allocation,  each firm minimizes its reduction, trade and terminal penalty cost till the system reaches a market equilibrium. Then, the regulator minimizes optimal social cost over possible allocations.   The stochastic model formalization goes as follows.
Consider a filtered probability space  $(\Omega, (\mathcal{F}_t)_{ 0\leq t\leq T}, \mathbb{P})$, in which the filtration is the augmented Brownian filtration generated by a standard $N+1$ dimensional Brownian motion $(\widetilde{W}^0,\widetilde{W}^1,\ldots, \widetilde{W}^N)$. Fix further correlation factors $(k_1, \ldots, k_N )$ and let
\begin{equation}\label{Wiener}
 W_t^i =  \sqrt{1-k_i^2}  \widetilde{W}_t^i + k_i  \widetilde{W}_t^0
\end{equation}
In particular, the correlation between $W^i,W^j$ is $\rho_{ij}:=  k_i k_j $. The index notation in what follows is self explanatory, in the sense that $Y^j$ denotes a process, while $Y_j,y_j$ refer to constants or deterministic functions.
 The Business As Usual (BAU) cumulative emissions dynamics for firm $i$ are:
\begin{align}
E^i_t = \mu_i t + \sigma_i W^i_t, \quad 0 \leq t \leq T,
\end{align}
where $\mu_i$ and $\sigma_i>0$ are the average growth rate and standard deviation rate of their emission. Thus, the emission of firm $i$ is affected by its own idiosyncratic noise $d\tilde W^i$ and by the common economic business cycle $d\tilde W^0$. A positive shock induces an increase in emissions.

In the BAU case, the expected total emission over the period $T$ is $\E[E_T]  = N \bar \mu T$ where $E_t = \sum_{i=1}^N E^i_t$ and $N\bar \mu = \sum_{i=1}^N \mu_i$. The regulation wishes to reduce the expected emissions to $$\rho \, T N \bar \mu,\  \  0<\rho<1$$
 i.e. to achieve $1-\rho$ percent of reduction compared to BAU.  The regulator has several instruments at her disposal (taxes, quotas) but she wishes to implement a dynamic cap--and--trade system working in this way.  {At $t=0$ she opens for  each firm a bank account $X^i$ and  allocates  permits, summed up in the cumulative process $\tilde{A}^i$}. The $i$-th bank  follows the dynamics:
\begin{align}\label{eq:bank-emissions}
dX^i_t  =  \beta^i_t dt + d \tilde{A}_t^i- dE^{i,\alpha^i}_t, \quad
dE^{i,\alpha}  = \big( \mu_i - \alpha^i_t \big) dt + \sigma_i dW^i_t.
\end{align}

In the above, $\alpha^i$ is the abatement rate and $\beta^i$ is  the trading rate in the (liquid) allowances market.  When making the effort $\alpha^i_t$, the emissions $E^{i,\alpha^i}$ of firm $i$ increases at a rate $\mu_i - \alpha^i_t$. \\

 \noindent {\bf Assumptions}: Firms controls are square integrable wrt $d\mathbb{P}\otimes dt$, namely they  belong to    $\L:=\TL$. The   allocation  process    $\tilde{A}^i$ is a square integrable semimartingale, that is $\E[(\tilde{A}_t^i)^2]<\infty $ for all $t$, decomposable into square integrable finite variation part $F$ and  square integrable stochastic integral:
  $$ \tilde{A}_t^i = F_t^i + \sum_{j=0}^N \int_0^t {\tilde{b}^j_s}d\widetilde{W}^j_s  $$
  
 \bigskip 
 
  \emph{We do not require that the finite variation part $F$ of $\tilde{A}^i$ is absolutely continuous wrt $dt$.} Therefore, the
 regulator is free e.g. to allocate (credit or debit) permits at discrete instants, which can be fixed or stopping times, and/or at a rate $\tilde{a}$.  In addition,  the Lebesgue Decomposition Theorem allows the decomposition of $F$  into the sum of an absolutely continuous part with respect to the Lebesgue measure, and of singular part $\tilde{S}^i$. So,
 \begin{equation}\label{eq:struct-all}
 \tilde{A}^i_t= \tilde{S}_t^i +\int_0^t \tilde{a}^i_t dt   +  \sum_{j=0}^N \int_0^t {\tilde{b}^j_s}d\widetilde{W}^j_s
 \end{equation}
 Each of the three addenda can be null.  As already mentioned, the singular  $\tilde{S}^i$ can be e.g. a pure jump part. Namely, a weighted sum of Dirac deltas  along a targeted  (stochastic) time grid, in which the regulator provides/cancels allowances in a discrete way: $$\tilde{ S}_t^i=  k^i_1 \delta_{t_1}(t)+k^i_2 \delta_{t_2}(t) \ldots $$
  where the dates  $t_h$ form an increasing sequence of stopping times and  $k^i_{t_h} \in \mathcal{F}_{t_h}$.  This covers the case in which there is  an initial allocation.  In fact,
   $$X^i_0 = \tilde{S}^i_0  $$
  it is enough  take $t_1=0$ in the grid, and $\tilde{S}^i_0 \neq 0$.  Given an allowance scheme, the allowance bank $X^i$ of firm $i$ gives at all  $t$s the net position of the firm in terms of emissions, abatement, allowances trading and allowances endowment by the regulator.  A positive economic shock to the emissions induces a decrease in the bank accounts, while an increase in the allocations  makes the accounts grow. The bank  dynamics can be rewritten as
\begin{align}
dX^i_t  =  d{A}^i_t+ \Big( \alpha^i_t + \beta^i_t  \Big) dt  - \sigma_i dW^i_t.
\end{align}
in which $A$ is the net allocation process over the trend
$$A^i_t =\tilde{A}^i_t -\mu_i t$$
Since our results on  the optimal regulation  depend only on $A^i$, \emph{the trend rate  $\mu_i$ does not necessarily have to be a constant.} Here, we set it constant for ease of presentation.
 For future use, note that
  $$ \E\big[ A_T^i \big ] = \E\big[ \tilde{S}^i_T \big] + \E\Big[\int_0^T (\tilde{a}^i-\mu_i) dt \Big]$$

For a given market price of permits $P$ and a given (net) allowance scheme ${A}^i$, the firm $i$ aims to  solve its cost minimization problem:
\begin{align} \label{J}
\inf_{\alpha^i, \beta^i} J^i(\alpha^i,\beta^i) :=  \inf_{\alpha^i, \beta^i} \E\Big[  \int_0^T   \Big( c_i(\alpha^i_t) + P_t \beta^i_t + \frac{1}{2 \nu} (\beta_t^i)^2 \Big) dt  + \lambda (X^i_T)^2 \Big].
\end{align}

In the objective function $J^i$ of firm $i$, the abatement cost function $c_i$ is supposed to be quadratic  $$c_i(\alpha) = h_i \alpha + \frac{1}{2 \eta_i} \alpha^2\  \ h_i,\eta_i>0 $$
  to consider both the linear and the adjustment costs, the latter proportional to  the square of the abatement rate. This choice is in line with the literature of carbon emission reduction (see Gollier (2020) \cite{Gollier20} and reference therein). Further, the linear-quadratic form  captures  the non decreasing feature of  marginal abatement cost. From an investment point of view, it means that there is some irreversibility in the abatement decision. The higher the values of $\eta_i$, the higher the flexibility of the abatement process and thus the higher the reversibility of the decision. \\
  \indent About trading costs, we take into account a price impact effect as in the original Kyle (1985) model \cite{Kyle85}, with constant market depth parameter ${\nu}>0$.  The term $\lambda (X^i_T)^2$, with  $\lambda>0$ equal for all firms,  is the terminal monetary penalty on the bank accounts set by the regulator.  The firm is going to pay  both if its bank is above or below the compliance zero level. It is a regularized version of the actual terminal (cap) penalty function, which is zero if the firm is compliant and linear otherwise.

The objective of the regulator is to design dynamic allocation schemes $\mathbf{{A}} = ({A}^1, \ldots {A}^N)$  to reduce expected emission, while minimising social cost
\begin{align} \label{eq:R}
& \inf_{\ \mathbf{{A}}\ } \E\Big[  \sum_{i=1}^N \int_0^T \big( c_i(\alpha^i_t) + \beta^i_t P_t + \frac12 \frac{(\beta^i_t)^2}{\nu} \big) dt + \lambda (X^i_T)^2 \Big], \quad  \E\Big[ \sum_{i=1}^N E^{i,\alpha^i}_T  \Big]  = \rho\, T N \bar \mu.
\end{align}
when the firms behave optimally and are at equilibrium, namely when $\sum_i \beta_t^i =0$.

So, the  problem  regulator/firms  falls in the category of dynamic stochastic Stackelberg games since the regulator  aims at minimizing social cost from optimal firms reduction and trade. Thus, she acts as a Leader while firms act as Followers (see Bensoussan et. al. (2014) \cite{Bensoussan14} for a survey).

 \vspace{5mm}

\section{The firm optimal response}
\label{sec:firm}

\subsection{General result}
\label{ssec:genoptresp}

The focus here is on the single firm cost minimization, for a given exogenous allowances price $P$ and a net allocation scheme $\tilde{A}^i$.  We define
\begin{align}\label{aAg}
  \quad M^i_t := \E_t \Big[ A^i_T\Big],  \quad  R^i_t := \E_t\Big[ A^i_T- A^i_t\Big], \quad g_i(t) := \frac{2\lambda \eta_i}{1+2\lambda(\eta_i+\nu)(T-t)}\end{align}
 in which
 \begin{itemize}
   \item $M^i$ is the martingale  closed by  $A^i_T$. It gives, at time $t$, the (conditional) expectation of the firm cumulative (net) endowment $A_T$. \emph{Different intertemporal allocations, with the same cumulative value on the regulatory horizon $[0,T]$ give rise to the same $M^i$}.  From the definition of $A^i, M^i$,
       $$M^i_0= \E\big[ A^i_T \big] = \E\big[ \tilde A^i_T \big] -\mu_i T$$

   \item  the process $ R^i$ gives, at time $t$, the conditional expectation of  the \emph{residual} net allocation on $[t,T]$.
 \end{itemize}

As anticipated in the Introduction, the firm minimization problem can be tackled by variational methods. In our case, the functional $J^i: \L\times \L \rightarrow \mathbb{R}$ is linear-quadratic:
$$ J^i(\alpha^i,\beta^i) =  \E\Big[  \int_0^T   \Big( c_i(\alpha^i_t) + P_t \beta^i_t + \frac{1}{2 \nu} (\beta^i_t)^2 \Big) dt  + \lambda (X^i_T)^2 \Big],$$  strictly convex and smooth. The optimal solution  is unique and can be obtained by annihilating the stochastic gradient.  \\
\indent By the Riesz Representation Theorem, any  linear form on a Hilbert space  can be represented by an element of the space itself. Therefore, the differential of $J^i$ can be represented by the gradient, which belongs to $\L\times \L$.   The gradient is then a couple of square integrable, adapted processes on which we are going to write  the first order conditions (FOC) in the proof of the next Theorem.

\begin{Theorem} \label{Theo:firm}
For a given cumulative, net allocation scheme $A^i$ and for a given allowances price $P$, the i-th firm  cost minimization problem \eqref{J} has a unique, explicit solution $(\hat \a^i,\hat \b^i)$ in $\L\times \L$. \\
\noindent {\rm (i)} The optimal abatement $\hat \a^i$ is the solution of the following SDE
\begin{align}\label{alphaopt}
d \hat\a^i_t &   = - g_i(t) \Big(    d (M^i_t-\sigma_i W^i_t) + d \E_t\Big[ \int_0^T \nu  (h_i-P_s) ds \Big]  \Big),  \\
\hat \a^i_0  & = - g_i(0) \Big( \frac{1}{2\lambda}  h_i + M^i_0  + \E \Big[ \int_0^T \nu  (h_i-P_t ) dt \Big] \Big)
\end{align}
and is therefore a martingale. \\
\noindent {\rm (ii)} The optimal trade $\hat\b^i$ is
\begin{align}\label{eq:betaopt}
 \hat \b^i_t =  \nu \Big( h_i + \frac{\hat \alpha^i_t }{\eta_i}  - P_t \Big).
 \end{align}\\
\noindent {\rm (iii)} Both optimal controls can be rewritten in feedback form in function of the bank state $\hat X^i_t$
\begin{align}\label{optimals}
 \quad \hat \alpha^i_t = \hat \a^i_t(\hat X^i_t) & = - g_i(t) \Big( \frac{h_i}{2 \lambda} + \hat X^i_t +   R^i_t +  \E_t\Big[ \int_t^T \nu (h_i-P_s) ds \Big] \Big) \\
 \hat \b^i_t = \hat \b^i_t(\hat X^i_t) & =  \nu \Big( h_i + \frac{\hat\a^i_t (\hat X^i_t)}{\eta_i} - P_t \Big) \label{beta}
\end{align}
in which the expected residual allocation process $R^i$ appears in place of $M^i$.
\end{Theorem}
\begin{proof}
The proof is given in Section~\ref{app:firm}.  Here, we just anticipate the FOCs written on the gradient of $J$, as they will be explicitly referred to in the rest of the paper. They are
\begin{align}
h_i + \frac{\a^i_t}{\eta_i} + 2\lambda \E_t\big[X^i_T\big] & =0, \label{eq:alpha-opt}\\
 P_t +\frac{\b^i_t}{\nu} + 2\lambda \E_t\big[X^i_T\big] &=0 \label{eq:beta-opt}
 \end{align}
\end{proof}

 Let us comment on these findings.
\begin{enumerate}[\upshape (i)]
\item The FOCs can be written
\begin{align*}
c'(\alpha^i_t) = - 2\lambda \E_t\big[X^i_T\big], \quad \b^i_t = \nu\big( c'(\alpha^i_t) - P_t\big).
 \end{align*}
The marginal abatement cost is equal to the (C.E. of the) marginal penalty, and so is the marginal cost of trading, $P_t + \beta^i_t/\nu$. {Consistently with economic intuition, the firm buys (resp. sells) if its marginal abatement cost is higher (resp. lower) than the market price.  }

\item The optimal  abatement  $\hat \a^i $ is a martingale. In fact, it is a stochastic integral, with a bounded, deterministic integrand $-g_i$, of \emph{three explanatory  martingales:} $M$, namely the C.E. of  cumulative net allocation over $[0,T]$; the conditional expectation of the  integrated price; and  the emission noise $\sigma_i W^i$.  A fortiori it is not the full intertemporal structure of the net allocation $A^i$ which matters, as it appears in  \eqref{eq:struct-all}. The key quantity here is $M$.\\
    \indent The firm compares the dynamics of the expectation of what will be given during the whole regulatory period, $A^i_T$, to noise and to integrated price, and then makes the decision on effort. If there is a positive economic shock $dW^i_t$ everything else being equal, the firm  effort increases. It decreases if $dM^i_t$ is positive, i.e. if the firm anticipates an increase in total expected net allocation.\\
    \indent The integrand $g_i$ in the martingale representation for $\hat\a^i$ depends on the following parameters: the individual firm adjustment cost of abatement $\eta_i$,  the common  penalty coefficient $\lambda$ and market depth coefficient $\nu$. \\
     When written in feedback form, $\hat \a^i$ depends only on the state $\hat X^i$, on the C.E. of net residual allocation $R^i$ at time $t$, $ R^i_t = \E_{t}[A^i_T-A^i_t]$, and on the C.E. of the (residual) integrated price.\\
     Finally, we remark that martingality of $\hat \a^i$ would be preserved if the abatement first order cost $h_i$ became a   martingale.

\item The optimal trade $\hat \b^i$ is not a martingale, unless $P$ is a martingale as well. This will occur at equilibrium (see Section~\ref{sec:market}).
    \end{enumerate}

\subsection{Market without frictions}\label{Sect:no impact}
 Absence of market frictions is a common assumption in the literature (see Kollenberg and Taschini (2016) \cite{Kollenberg16} or Carmona et. al. (2010) \cite{Carmona10} and the references within). TO better compare with this case, let us solve the firm  optimisation problem  when the market has infinite depth, $\nu =\infty$. The problem becomes
\begin{align} \label{Jinfty}
\inf_{\alpha^i, \beta^i} \widetilde{J}(\alpha^i,\beta^i) := \inf_{\alpha^i, \beta^i} \E\Big[  \int_0^T \left (  h_i \a^i_t + \frac{(\a_t^i)^2}{2\eta_i}    + P_t \beta^i_t  \right ) dt  + \lambda (X^i_T)^2 \Big]
\end{align}
If the optimizers exist,   we cannot expect that $\biopt$ will be unique.  The objective function in fact loses strict convexity in the $\b$ argument.  The quadratic terminal penalty however involves the cumulative trade $B^i_T= \int_0^T \b^i_tdt$, for which uniqueness will be obtained. 

\begin{Proposition}  \label{prop:firm-no-nu}

Problem \eqref{Jinfty} admits a solution if and only if $P$ is a martingale.
 In case $P$ is a martingale,
the abatement effort of firm $i$ is unique and given by:
\begin{align}
\hat \alpha^i_t & = \eta_i \big( P_t - h_i \big).
\end{align}
 The optimal trade rate is not unique. Any $\b^i \in \L$ satisfying
\begin{equation} \label{Biopt-inf}
\int_0^T \b^i_t dt = \hat B^i_T
\end{equation}
is optimal, where $\hat B^i $ is the $\L$ martingale satisfying the Cauchy problem
\begin{align} \label{Biopt-inf2}
& d \hat B^i_t = - \left(\frac{1+ 2\lambda(T-t)\eta_i}{2\lambda} dP_t  + d(M_t^{i} - \sigma_i W^i_t) \right), \quad
\hat B_0^i =   -\left( \hat P_0\frac{(1- 2\lambda \eta_i T)}{2\lambda} + M_0^i + \eta_i h_i T \right).
\end{align}

\end{Proposition}

 The complete proof follows the same lines as the previous Theorem \ref{Theo:firm}. We briefly   highlight the main differences.   The resulting FOCs are:
  \begin{align}
& c'(\alpha^i_t) + 2\lambda \E_t \big[ X^i_T \big] =0, \label{alpha-noimpact} \\
& P_t  + 2\lambda \E_t[X^i_T] = 0. \label{FOC-no impact}
\end{align}
If $P$ is not a martingale, there are no stationary points and thus no minimizers. When $P$ is a martingale, as economic intuition suggests,  each firm equates the marginal cost of abatement  to the market price $P_t$. Also,  the market price is equal to the conditional expectation of the marginal penalty. In Theorem \ref{Theo:firm}, frictions introduce deviation from these equalities. With finite $\nu$ in fact, we saw that  the marginal cost of abatement equals  the marginal cost of trading, $P_t + \beta^i_t/\nu$ as from \eqref{eq:alpha-opt}, \eqref{eq:beta-opt}. Same holds for the relation between the marginal cost of trading and the marginal penalty. \\
\indent Further, the FOC equations here
do not involve $\b^i$ directly, but only the martingale $B^i$ generated by the total trade:
  $$
  B_t^i : = \E_t\left [\int_0^T\b^i_tdt \right ].
  $$
This is the main novelty, now the  optimisation problem is strictly convex only in the total trade $B_T^i$. Therefore the optimal $\hat B^i_T$ and the  generated martingale $\hat B$  are unique. And, in fact, such martingale  is found by replacing in relation~\eqref{FOC-no impact}, $\hat \alpha$ by its expression as a function of $P$:
    \begin{equation} \label{Pdyn-noimpact}
     P_t +  2\lambda  M^i_t  + 2\lambda \E_t \Big[ \int_0^T \eta_i(h_i -   P_t) dt  \Big] + 2\lambda B^i_t - 2 \lambda \sigma_i W^i_t =0
    \end{equation}
An  application of Lemma \ref{CE-of-Mart-dyn}, together with evaluation at $t=0$, gives
\begin{align} \label{Biopt-inf3}
d \hat B^i_t & = -\left ( \frac{1+ 2\lambda(T-t)\eta_i}{2\lambda} dP_t  + dM_t^{i} - \sigma_i dW^i_t \right ), \\
\hat B_0^i & = -\left( \hat P_0\frac{(1- 2\lambda \eta_i T)}{2\lambda} + M_0^i + \eta_i h_i T \right)
\end{align}
Therefore, any $\b^i$ satisfying
\begin{equation*}
\int_0^T \b^i_t dt = \hat B^i_T
\end{equation*}
is optimal.

\section{Market equilibrium}
\label{sec:market}

We are now ready to tackle the equilibrium problem of the system of  $N$ firms. Recall that the  noises $W^i$ in the firms activity   have a quite general dependence structure, as described at the beginning of Section \ref{sec:firm}. Fix a net allocation policy of the regulator  $\mathbf{A}=(A^1, \ldots , A^N) \in (\L)^N.$  Define the positive, deterministic function
\begin{align}
 \pi_i(t) & := \frac{g_i(t) }{\eta_i  } \Big(1 -    \frac{\nu (T-t)}{N} \sum_{k=1}^N \frac{g_k(t)}{\eta_k}  \Big)^{-1},
  \end{align}
 where $g_i$ is defined in \eqref{aAg}. If firms share the same $\eta_i$, all the functions $\pi_i$ are equal. Market equilibrium consists in finding  a  price $\hat P$ that satisfies the market clearing condition:
\begin{align}
\sum_{i=1}^N \hat \b_t^i( \hat P) = 0, \quad \forall t \in [0, T].
\end{align}
in which the $\hat \b^i$ are given from the system \eqref{eq:betaopt}. The price $\hat P$ is then  called an \emph{equilibrium price}.
The market equilibrium is described in the following Theorem.

\begin{Theorem} \label{Theo:equilibrium}
For a given net cumulative allocation $\mathbf{A}$,
\begin{enumerate}[\upshape (i)]
\item The equilibrium price $\hat P$ is the unique  solution to the Cauchy problem:
\begin{align}
\label{P-EQPRICE}
d\hat P_t  =  - \frac{1}{N} \sum_ {i=1}^N   \pi_i(t) \big( dM_t^{i}  - \sigma_i dW_t^i \big),   \quad
\hat P_0   =   \frac1N   \sum_{i=1}^N  \pi_i(0) \Big( \eta_i h_i T - M_0^i\Big).
\end{align}
The price $\hat P$ is therefore a martingale.
\item The equilibrium $\hat P$ can be written in feedback form as
\begin{align}\label{price-and-pi}
\hat P_t =  \frac1N  \sum_{i=1}^N \pi_i(t) \Big(\eta_i h_i {(T-t)}- (\hat X^i_t +   R^i_t)  \Big).
\end{align}
\item The optimal controls $\hat \a^i,\hat \b^i$  are the unique  solutions of the next Cauchy problems:
\begin{align} \label{aiopt}
d\aiopt_t & = -g_i(t)  \Big[  d(M_t^{i}  - \sigma_i  W_t^i) -  \nu(T-t) d\hat P_t \Big], \quad
\aiopt_0  =  -g_i(0) \Big[ h_i \Big(\frac{1}{2 \gl} +\nu T \Big) + M_0^i  -\nu T \hat P_0 \Big], \\
d\biopt_t & =    \nu d (\hat P_t - h_i - \frac{1}{\eta_i}  \hat \alpha^i_t),  \quad \biopt_0  =  \nu \Big( h_i + \frac{\aiopt_0}{\eta_i} - \hat P_0 \Big).
\end{align}
\item In feedback form,
\begin{align}\label{eq:Rem-FRICTIONS}
\hat \alpha^i_t(\hat X^i_t, R^i_t) & =  g_i(t) \Big[  \nu (T-t)( \hat P_t - h_i)  -  \Big(\frac{h_i}{2\lambda} +  \hat X^i_t +   R^i_t \Big)  \Big],  \quad  \hat \b^i_t(\hat X^i_t, R^i_t)  = \nu (\hat P_t - \frac{\hat\a^i_t}{\eta_i} -h_i ).
\end{align}
\end{enumerate}
\end{Theorem}
\begin{proof} See appendix~\ref{app:equilibrium}.
\end{proof}

Let us comment on the above results.
\begin{enumerate}[\upshape (i)]
\item The explanatory processes in \eqref{P-EQPRICE} for the equilibrium price $\hat P$ are the  martingales  $(M^i -\sigma_i W^i) $,  $i=1, \ldots, N$. We observe that if all these martingales experience a positive shock, the price decreases. In short, if the regulator promises to all firms more  (resp. less) future total net allocation than the effect of their economic shock, the price decreases (resp. increases).

\item When the adjustment costs $\eta_i$ are equal,   the deterministic coefficients $\pi_i$ become equal and can be factorized out in \eqref{price-and-pi}. The equilibrium price $\hat P$ in this case depends then \emph{only on the aggregrate quantities } \begin{equation} \label{eq:aggr-quant}
    Z_t := \sum_{i=1}^N (M^i_t - \sigma_i W^i_t )\quad \text{and} \quad N \bar h := \sum_{i=1}^N h_i
    \end{equation}
   \item The optimal efforts $\aiopt$ are obtained from \eqref{alphaopt} exploiting the martingality of $\hat P$. The trade $\biopt$ keeps  its structure.
\end{enumerate}

\vspace{5mm}

 When there are no market frictions, we have seen in Section \ref{Sect:no impact} that the optimal trade is non unique. However, there is a unique (martingale) equilibrium price $\hat P$, and consequently a unique abatement effort $\hat \a^i, i=1, \ldots N$. The equilibrium price $\hat P$ depends only on the aggregate quantities as in \eqref{eq:aggr-quant}, plus the average of adjustment cost coefficient $\bar \eta$.  If $(\mathbf{\hat \a }, \mathbf{\hat \beta }, \hat P)$ denotes an equilibrium triplet, the next Proposition sums up the results in this particular framework. We omit the proof, since  it follows from combining Theorem \ref{Theo:equilibrium} (sending $\nu$ to infinity) and Proposition \ref{prop:firm-no-nu}.

\begin{Proposition}   \label{prop:eqmarketnu0}
When there are no market frictions,
\begin{enumerate}[\upshape (i)]
\item The equilibrium price dynamics become
\begin{align}
  & d\hat P_t  =  - f(t)  \Big(  d\bar M_t -   d\bar W_t \Big), \label{eq:P0-0-noimpact}
  \quad
  \hat P_0   = f(0) \Big( T \bar H  - \bar M_0   \Big), \\
& \text{with}  \quad \bar M_t := \frac1N \sum_{i=1}^N M^i_t, \quad   \bar W_t := \frac1N \sum_{i=1}^N \sigma_i W^i_t, \quad f(t) := \frac{2\lambda}{1+2\lambda\bar\eta(T-t)}. \nonumber
\end{align}
in which $\bar H := \frac1N \sum_{i=1}^N \eta_i h_i,$ and $ \bar \eta := \frac1N \sum_{i=1}^N \eta_i.$
Its expression  in closed-loop is given by
\begin{align}\label{eq:Phatnu0}
 \hat P_t &  = f(t)\Big( (T-t) \bar H -  \bar X_t - \bar R_t   \Big)
\end{align}
where $\bar X$ denotes the average bank account process. 
\item The abatement effort of firm $i$ is unique and given by:
\begin{align}\label{eq:alpha-no-nu}
\hat \alpha^i_t & = \eta_i \big( \hat P_t - h_i \big).
\end{align}
\item The trading rates are not unique. Any $\b^i \in \L$ satisfying
\begin{equation} \label{Biopt-inf}
\int_0^T \b^i_t dt = \hat B^i_T
\end{equation}
is optimal, where $\hat B^i_t$ satisfies the Cauchy problem
\begin{align} \label{Biopt-infB}
 d\hat B^i_t  = -\left ( \frac{1+ 2\lambda(T-t)\eta_i}{2\lambda} dP_t  + dM_t^{i} - \sigma_i dW^i_t \right ), \hspace{0mm}
\hat B_0^i  = -\left( \hat P_0\frac{(1- 2\lambda \eta_i T)}{2\lambda} + M_0^i + \eta_i h_i T \right).
\end{align}
\end{enumerate}
\end{Proposition}

\section{Optimal dynamic regulation}
\label{sec:optimalregulation}

Market frictions are small compared to the cost of abatement required to achieve the carbon emission reduction targeted by the European Union. As documented by Frino et. al. (2010), the carbon market quality has been constantly increasing with tick size decreasing from 5~c\euro/t to 1~c\euro/t and a bid--ask spread of 5~c\euro/t as of 2008. Nowadays, the value of the bid--ask on the December contract is around 2 c\euro/t for a quoted carbon price around 30~\euro/t\footnote{Source: Thomson-Reuters Refinitiv quotations of the EU EUA December contract.}, which makes a transaction cost less than~0.06\%. Hence, we neglect them in the regulator's problem  and assume hereafter that  \emph{there are no market frictions.}

\subsection{Main result}
\label{ssec:mainres}

We address now the optimisation problem~\eqref{eq:R} of the regulator when the market is at equilibrium.    The regulator faces:
\begin{align}\label{eq:Reg-OF}
\inf_{\mathbf{A}} C(\mathbf{A}) : = \E\Big[ \sum_{i=1}^N \int_0^T \Big( h_i \hat \a^i_t + \frac{(\hat \alpha^i_t)^2}{2\eta_i} \Big) dt + \lambda (\hat X^i_T)^2\Big], \quad \E \big[ E^{\hat \alpha} _T \big] =  \rho\, T N \bar \mu.
 \end{align}
in which $\hat \a =(\hat \a^1, \ldots, \hat \a^N)$ is the optimal effort of the system given the allocation $\mathbf{A}$, while $E^{\hat \a}$ is the system emission under this effort. 
Using Proposition~\ref{prop:eqmarketnu0}~(i) and (ii),
\begin{align*}
\E \big[ E^{\hat \alpha} _T \big] = N T \big( \bar \mu - \bar \eta \hat P_0 + \bar H\big).
\end{align*}
Hence, the reduction constraint on the expected emissions amounts to \emph{a constraint on the equilibrium price}
\begin{align}\label{eq:Phat0}
 \hat P_0 & =  \frac{1}{\bar \eta} \big( \bar H + (1-\rho) \bar \mu \big).
\end{align}

The relation~\eqref{eq:Phat0} gives the necessary average carbon price required to emerge from the market if the regulator wishes to achieve a carbon emission by a factor $1-\rho$. This average price captures all the features of the system in a simple way: it is proportional to the {\em inflexibility} $1/\bar \eta$ of the system, to the average abatement cost $\bar H$, to the growth rate of the emission $\bar \mu$ and to the ambition of reduction $\rho$.

Equating the constraint on $\hat P_0$ in \eqref{eq:Phat0} with the expression~\eqref{eq:P0-0-noimpact} of $\hat P_0$  obtained in Proposition \ref{prop:eqmarketnu0}~(i), we get
\begin{align}\label{eq:x0M0}
\bar{M}_0 = - \frac{1}{2\lambda \bar \eta} \Big[ \bar H + \big( 1 + 2\lambda \bar \eta T \big) (1-\rho) \bar \mu \Big] =: \ell(\rho) <0.
\end{align}

Thus, \emph{the regulator withdraws allowances on average as $\bar M_0$ is negative}. This holds regardless of the intertemporal allocation processes $A^i$ from the (detrended) equation   \eqref{eq:struct-all}, as the relevant processes are the $M^i$.  If e.g. firms receive allowances in the beginning, so that the banks $X^i$ satisfy $\sum_i X^i_0 = \sum_i  {A}_0^i = \sum_i \tilde{S}^i_0 >0$,  then in $(0,T]$ the regulator on average will withdraw permits.\\
\indent Conversely, if initially firms are given $ \bar X_0 := \sum_i X^i_0/N  <  \ell(\rho)$
 then the regulator will on average credit back permits in $(0,T]$. In fact, using the decomposition of $A^i$,
 \begin{align*}
    \bar M_0 = \E[\bar{A}_T] =    \bar X_0   + \frac{1}{N}\sum_{i=1}^N \E \Big[  \tilde{S}^i_T- \tilde{S}^i_0 +   \int_0^T a^i_t dt \Big] =\ell(\rho)
    \end{align*}
and  the second addendum must then be positive.  By the Predictable Representation Property of the Brownian filtration  generated by the $N+1$ noises $\widetilde{W}^j, j=0, \ldots, N$ (see \eqref{Wiener}), each  $M^i$ can be written  as
 $$ M^i_t =  M_0^i + \int_0^t \sum_{j=0}^N  \gik_s d\widetilde{W}^j_s  $$
with $\gik \in \L$, $j=0, \ldots N$.  If $\gamma^i = (\gamma^{i,0}, \ldots, \gamma^{i,N}) $ denotes the integrands vector for firm $i$, the  problem can  be re-parametrized on controls $ \mathbf{ M }\equiv ( (M_0^1, \gamma^1), \ldots, (M^N_0, \gamma^N)) $. With this  formulation, we state   our main result.

\begin{Theorem} \label{social-cost-min} The social cost minimisation problem
\begin{align} \label{R}
  \inf_{\mathbf{A}} C (\mathbf{A})  =  \inf_{\mathbf{M}} \E\Big[ \sum_{i=1}^N \int_0^T \Big( h_i \hat \a^i_t + \frac{(\hat \alpha^i_t)^2}{2\eta_i} \Big) dt + \lambda (\hat X^i_T)^2\Big],   \quad   \bar{M}_0 =  \ell(\rho),
\end{align}
has the following solutions structure:
\begin{enumerate}[\upshape (1)]
\item  Optimizers $\hat \gamma^i$, $i=1, \ldots, N$ annihilate the volatility of the price.  The vectors $\hat \gamma^i$ are non unique. One set of optimizers is obtained by  tracking the volatility of each firm separately, thus allocating to the i-th firm  exactly the systemic ($j=0$) and idiosyncratic components ($j=i$):
\begin{align*}
  \hat \gamma^{i,0}_t  = \sigma_i k_i, \quad
  \hat \gamma^{i,i}_t   = \sigma_i \sqrt{1-k_i^2}, \quad \text{ and }
     \hat \gamma^{i,j}_t  = 0  \quad \text{ for } j \neq 0 \text{ and } j\neq i.
\end{align*}
The optimal martingales $\hat M^i_t$ become
$$ \hat M^i_t = \hat M_0^i +  \sigma_i  W^i_t, \quad  i =1\ldots N$$

\item Expected optimal allocations $\hat M_0^i$ are also non unique. The regulator is free to allocate permits as long as the expectation of the total number of permits   satisfies the constraint $\sum_{i=0}^N \hat M_0^i = N \ell (\rho)<0$. An example is the equal assignment in expectation: 
    $$ \hat M^0_i = {\ell (\rho)} \ \ \text{ for all } i $$

\item Optimal allocations $\mathbf{A}$ are, as a consequence, non unique.   If the regulator chooses the firm-by-firm volatility tracking and equal assignment in expectation as in item 1) and 2)  above, there is an optimal set of net martingale allocations:

$$A_t^i   =  {\ell (\rho)} +  \sigma_i W^i_t =\hat M^i_t $$

\item As the price has zero volatility, it is  constant:
      \begin{align} \label{eq:price-opt-dyn}
\hat P_ t=  \hat P_0 =   \frac{1}{\bar \eta} \big( \bar H + (1-\rho) \bar \mu \big)
 \end{align}
\item Firms optimal abatements are unique and constant \begin{align*}
\hat \alpha^i_t =  \hat \alpha^i_0 =  \eta_i (\hat P_0 - h_i ).
 \end{align*}
 \item Firms optimal trading rates are non unique. Any $\hat \beta^i$ satisfying \eqref{Biopt-inf} is optimal. If the regulator chooses the firm-by-firm individual tracking in item 1) above,  the dynamics of the associated $\hat B^i$ are null. If in addition, there is equal endowment in expectation then an optimal solution is trading at a constant rate
      $$ \hat{\beta_t^i} = \frac{\hat B^i_0}{T} =\frac{1}{T} \left( \frac{(1+2\lambda \eta_i T)}{2\lambda} \hat P_0 +  \ell(\rho) - \eta_i h_i T \right )$$
\item The minimum social cost is
      \begin{equation}\label{eq:social-cost}
       \hat C = \frac{N}{4\lambda}  ( 1+ 2\lambda \bar \eta T) \hat P_0^2  -  \frac{1}{2} T \sum_{i=1}^N \eta_i h_i^2
       \end{equation}

\end{enumerate}

\end{Theorem}

\begin{proof}
 The cost function in \eqref{R} is convex and differentiable in $M^i_0, \gamma^i $ for all $i$.
The Lagrangian is
$$ L (M,\xi)= \sum_{i=1}^N \E\Big[  \int_0^T \Big( h_i \hat \a^i_t + \frac{(\hat \alpha^i_t)^2}{2\eta_i} \Big) dt + \lambda (\hat X^i_T)^2\Big]  + \xi \left (  \sum_{i=1}^N  M^i_0   \, - N \ell(\rho)\right)$$
in which the optimal controls are from Proposition~\ref{prop:eqmarketnu0}.
   From \eqref{FOC-no impact}, $\hat P_T = -2\lambda \hat X^i_T$ for all firms $i$. Substituting  the optimal firm abatement controls with  $\hat P $  as from \eqref{alpha-noimpact}, the optimal cost for the $i^{\rm th}$ firm is
$$  \E\Big [\int_0^T   h_i \eta_i(\hat P_t - h_i)  + \frac{1}{2\eta_i}  (\eta_i(\hat P_t - h_i))^2 dt + \lambda  \frac{1}{4\lambda^2} (\hat P_T)^2 \Big] $$
 By the martingale property of $\hat P$,   the Lagrangian becomes
$$L(M, \xi)= -\frac{1}{2} T \sum_{i=1}^N h_i^2\eta_i  + \frac{1}{2} \sum_{i=1}^N \eta_i  \E\Big[\int_0^T \hat P_t^2 dt \Big] + \frac{N}{4\lambda} \E\big[ \hat P_T^2 \big] +  \xi \left (  \sum_{i=1}^N  M^i_0  \, - N \ell(\rho)\right)$$
Also,
$$ \E\big[ \hat P^2_t\big] = \hat P^2_0 + 2 \E\big[ ( \hat P_t -\hat P_0) \big] + \E\big[  (\hat P_t -\hat P_0)^2\big ] =  \hat P_0^2 + \E\big[ \langle \hat P\rangle_t\big]$$
Finally,
$$ L(M, \xi)= -\frac{1}{2} T \sum_{i=1}^N h_i^2 \eta_i + \frac{N}{4\lambda} ( 2\lambda T \bar \eta + 1) \hat P_0^2  + N \frac{\bar \eta}{2} \E\Big[\int_0^T  \langle \hat P\rangle_t dt \Big] + \frac{N}{4\lambda} \E\big[\langle \hat P\rangle_T \big] +  \xi \left (  \sum_{i=1}^N  M^i_0  \, - N \ell(\rho)\right)$$

Optimality conditions:

\begin{itemize}
  \item $\nabla_{M_0} L $. We have to impose
\begin{eqnarray*}
 \frac{\partial L}{\partial M_0^i}  =\xi -  \frac{N}{4\lambda} ( 2\lambda T \bar \eta + 1) \frac{f(0)}{N} = \xi -  \frac{\hat P_0}{2}   = 0  \ \ \ i= 1 \ldots N
 \end{eqnarray*}

\item $\nabla_{\gamma} L $. The matrix process $\gamma$ in the representation of $M$  is involved only in the dynamics of $\hat P$. In the Lagrangian, it thus enters only in the quadratic variation.  This implies that the minimum wrt $\gamma$ is attained when the regulator annihilates the quadratic variation, i.e.  the volatility of $\hat P$. From Proposition \ref{prop:eqmarketnu0}, we just need to impose that
     $$ \langle \bar M -\bar W\rangle =0 , $$
    in which $\bar W = \frac{1}{N}\sum_{i=1}^N \sigma_i W^i$. Using \eqref{Wiener}, and recalling $M^i_t =  M_0^i + \int_0^t \sum_{j=0}^N  \gamma^{i,j}_s d\widetilde{W}^j_s  $, the previous equation can be rewritten as
       \begin{align*}
        \langle \sum_{i=1}^N \int_0^\cdot \sum_{j=0}^N  \gamma^{i,j}_s d\widetilde{W}^j_s- \sum_{i=1}^N \sigma_i(\sqrt{1-k_i^2}\widetilde{W}^i+k_i\widetilde{W}^0) \rangle =0.
        \end{align*}
   The above boils down to the system:
   \begin{align}\label{sys-opt-gamma}
   \sum_{i=1}^N(\gamma^{i,0}_s - \sigma_ik_i)  =0, \ j=0,  \quad
    \sum_{i=1}^N\gamma^{i,j}_s -\sigma_j \sqrt{1-k_j^2}  =0 , \ j=1, \ldots, N
   \end{align}
 Namely, the regulator allocates  on the $j^{\rm th}$ Brownian motion $\widetilde W^j$ the aggregate volatility the system has in that shock. Thus, the net effect is that the regulator  kills the exposure to shocks in  the whole system.   Clearly, a particular solution is to kill exposure firm by firm, with $ \gamma^{i,0}_s - \sigma_ik_i=0 $, $\gamma^{j,j}_s -\sigma_j \sqrt{1-k_j^2}=0, $ and $\gamma_s^{i,j} =0$ otherwise as stated in item (1).

\item Feasibility is simply $$ \sum_i M^i_0 - N\ell(\rho) = 0$$
\item Price will then be unique, the positive constant
     $$ \hat P_0 = \frac{2\lambda }{1+2\lambda \bar \eta T} (T\bar H - \ell (\rho)) = \frac{1}{\bar \eta} \big( \bar H + (1-\rho) \bar \mu \big)$$
\item The optimal efforts $\hat \alpha^i$ are unique and constant, as per \eqref{eq:alpha-no-nu}.
\item The optimal C.E. of total trade $B^i$ are non unique,  because they depend on the C.E of the \emph{individual} optimal total allocation  $\hat M^i$ as detailed in \eqref{Biopt-infB}. If the regulator tracks individual volatility, then $\hat M_t^i = \ell(\rho) + \sigma_i W_t^i$ and   $\hat B^i = const = \left( \frac{(1+2\lambda \eta_i T)}{2\lambda} \hat P_0 +  \ell(\rho) - \eta_i h_i T \right )$.
\item Finally, the expression of the minimum social cost easily follows from the above relations.
\end{itemize}\hfill$\Box$
\end{proof}

Hence, the dynamic regulation effect is a constant market price. However, this condition is not imposed  {\em a priori}, but it is a consequence of social cost minimisation. Indeed, the level of carbon emission reduction fixes the average level of required effort, and thus the average required price. But price fluctuations  induce variations of effort, which in turn produce irreversible cost because of the inflexibility of the system. In fact, social costs are increased by  price fluctuations. By annihilating price changes, the regulator avoids   firms expensive stop-and-go. \\
\indent In the optimal policy given  in Theorem \ref{social-cost-min},  one has
\begin{align*}
\tilde A^i_t = A^i_t + \mu_i t =  \ell(\rho) + \sigma_iW^i_t +  \mu_i t,
\end{align*}
i.e. the regulator provides an initial (negative) allocation and then, credits  the whole BAU emissions to the firms.   As a consequence, the  bank accounts have a deterministic, linear dynamic:
\begin{align*}
 \hat X^i_t =  \ell(\rho) + (\hat{\alpha}^i_0  + \hat{\b}^i_0) t \ \ \quad  i=1, \ldots, N.
\end{align*}

  In other optimal schemes,  the regulator eliminates all the economic uncertainty involved in the dynamics of the carbon emissions. Regulation does not necessarily kill  individual emission noises, but uncertainty is tackled as a whole - as soon as the optimality condition \eqref{sys-opt-gamma} is respected. \\
\indent  Although more complex to implement than a tax or a static initial endowment,  optimal dynamic policies offer a  powerful  tool in emissions control. In particular,    non-uniqueness of optimal dynamic policies is a key feature in our model. Indeed, because they are non-unique, the regulator can  achieve more   goals  using the same device. This can be obtained by e.g. adding more constraints to the optimal control problem of the regulator. An example could be an additional constraint on the financing of new technologies. We leave these developments for future research.

\begin{Remark}[When there are frictions]
The same procedure used in Theorem~\ref{social-cost-min} applies to the case where there are market frictions. In that case, Theorem~\ref{Theo:equilibrium} states that the optimal efforts and trading rates are martingales. Also, the terminal values of the bank accounts still are deterministic functions of $\hat P_T$, as can be easily deduced from \eqref{eq:alpha-opt} and \eqref{eq:Rem-FRICTIONS}.  Therefore the regulator's general problem ~\eqref{eq:R} also  amounts to annihilating the volatility of $\hat P$. This goal is uniquely achieved by individual tracking,  giving $M^i =M^i_0 + \sigma_i W^i$ to firm $i$.  The dynamic allocation proposed in Theorem~\ref{social-cost-min}, item (3) is now the unique optimal policy. Optimal efforts and optimal trading rates will also be unique.
\end{Remark}

\subsection{Extensions}
\label{ssec:extensions}

\paragraph{Cap-and-trade mechanism}
A cap-and-trade mechanism is often described by a terminal penalty in the form of a cap on emissions. This corresponds to a put on the bank, with strike equal to the maximum tolerated emission level $L$. We show that   the firm optimization problem  can be still be solved by the variational methodology. Suppose the cost function $K^i$ of the agent is
 $$ K^i(\alpha^i, \beta^i)= \E\Big[  \int_0^T   \Big( h_i \alpha^i_t + \frac{(\alpha_t^i)^2}{2\eta_i} - P_t \beta^i_t + \frac{1}{2 \nu} (\beta_t^i)^2 \Big) dt\Big ]+ \E[\lambda (L_i-X^i_T)^+]$$
where the cap $L$ is the minimum tolerated bank level at the end of the regulated period. After this threshold, the firm pays at maturity $ \lambda$ per ton exceeding the bank level. This formulation can be seen as continuous time version of the model in Carmona, Fehr an Hinz \cite{Carmona09}, with the additional features of price impact and quadratic abatement costs. In order to solve
$$ \min_{\a^i, \b^i} K^i(\alpha^i, \beta^i) $$
we  observe that the functional $K^i$ is  finite, strictly convex and coercive on $\L \times\L$ and so there exists a unique optimizer. $K^i$ is  also sub-differentiable. In fact, the pointwise (non adapted) subdifferentials of the put  expectation are:
   $$\widetilde{\lambda} = -(\lambda \1_{\{L- X^i_T > 0 \}} + \tau \1_{\{L- X^i_T = 0 \}}) $$
 in which  $0\leq \tau\leq \lambda $ is a   $\mathcal{F}_T$-measurable r.v.  Note that when  $\P(L- X^i_T = 0  )=0$ then  all  the subdifferentials at $X^i_T=L$ coincide almost surely, so we can safely choose $\tau = \lambda $. The put becomes differentiable. This happens e.g. when the allocation of the regulator does not perfectly track the bank noise. In fact, in this case  the distribution of $X^i_T$ has no atoms thanks to the presence of a diffusion term.   In the paper \cite{Carmona09},  the assumption (19) in Theorem 1 is a very similar condition on the net firm position, which ensures that the subdifferential of the put is in fact unique.
 To better compare with \cite{Carmona09}, assume the put is differentiable.  The FOCs become:
\begin{align}\label{differential-bis}
  \frac{\partial}{\partial \alpha^i}K^i &= h_i + \frac{\a^i_t}{\eta_i} - \lambda \E_t[\1_{\{L- X^i_T \geq 0 \}} ] =0, \\
   \frac{\partial}{\partial \b^i} K^i  &=  -P_t +\frac{\b^i_t}{\nu} - \lambda \E_t[\1_{\{L- X^i_T \geq 0 \}} ]=0
\end{align}
and the system admits a unique solution in $ \L \times \L$. At equilibrium,  by market clearing   the $\b$ term vanishes in the second FOC and  \emph{the equilibrium price becomes  the average of the conditional expectations of the marginal penalties.}  If, as in \cite{Carmona09} we neglect price impact, at equilibrium one gets the simplified relation
 $$ P_T = \lambda \1_{\{L- X^i_T \geq 0 \}} \ \ \  i=1, \ldots, N$$
This implies that the sets $\{L- X^i_T \geq 0 \}, i=1 \ldots N $,  coincide a.s. and
 $$   N L -\sum_{i=1}^N X^i_T \geq 0 \quad  \mathrm{ a.s. } \quad  \text{ iff } L -X^i_T\geq 0 \quad \mathrm{ a.s.} $$
 and therefore the price can be seen as  the marginal penalty of the aggregate put:
 $$P_T = - \lambda \1_{ \{  N L -\sum_i X^i_T \geq 0  \}}$$
 and equals  the marginal abatement costs.

 As in the cited \cite{Carmona09},  optimal strategies cannot be found in closed form.  Nevertheless, we observe that the market equilibrium price is still a martingale, depending on the allocation processes and the abatement efforts are a linear function of the price. As in  Theorem \ref{social-cost-min} then, optimal allocations  annihilate the volatility of the price.

\paragraph{Risk aversion taken into account}
Let us denote by $$ Y^i_T  = \int_0^T \left( h_i \a^i_t +  \frac{1}{2\eta_i}(\a_t^i)^2 -P_t\b^i_t  + \frac{(\b_t^i)^2}{2\nu}\right )dt +  \lambda (X_T^i)^2$$
the random  cost incurred in $[0,T]$ by a single firm.  Suppose the agent has a concave utility $U_i$ finite on $\mathbb{R}$,  smooth and with $U_i'>0$. Consider the expected utility functional
$$ (\a, \b)\in \L\times \L \rightarrow   \E[U_i(-Y_T)]   $$ 
 Assume that the functional is proper, namely there exists a couple $\a^0,\b^0$ in $\L$ such that the expected utility of the associated cost $Y^0$, $E[U_i(Y^0_T)]$, is finite. The agent seeks to solve
$$  \max_{\a, \b} \E[U_i(-Y_T)] $$

Proceeding heuristically, we differentiate under the integral sign and get the (non adapted) gradient 
  $$ - U_i'(-Y_T) DY $$
  in which $DY$ is the bivariate process:
  $$  DY_t = \left( \begin{array}{cc}
                      h_i + \frac{\a_t}{\eta_i} + 2\lambda \E_t[ X_T] , &  -P_t + \frac{\b_t}{\nu} -2\lambda \E_t[ X_T]
                    \end{array}
   \right)
   $$
Since $U_i' \neq 0$, then the FOC condition is
 $$ DY =0 \ \ dP\otimes dt \text{ a.e.} $$
 which is equivalent to  \eqref{differentialJ} and therefore the same we had in the risk neutral case.
 Thus, the solutions found in Theorem \ref{Theo:firm} are the candidate optimal couple.  We only need to check that the associated  cost $\hat Y_T$ has finite utility.  This is straightforward
 $$ -\infty < \E[U_i(\hat Y^0_T)] \leq  \E[U_i(\hat Y_T)]\stackrel{concavity}{\leq} U_i ( \E[ \hat Y_T]) <\infty  $$
 since $\hat Y_T$ here is integrable and $U_i$ is finite on $\mathbb{R}$.  Therefore, $(\hat \a, \hat \b)$ from Theorem \ref{Theo:firm} continues to be the optimal couple even in the presence of risk aversion.

\section{Comparison with existing policies}
\label{ssec:altpol}

We compare here the optimal dynamic allocation policy suggested in Theorem~\ref{social-cost-min} with  three alternative existing policies: the initial static allocation, the pure tax system,  and a Market Stability Reserve--like allocation mechanism. In each case, we compute the social cost to achieve the same expected total carbon emissions reduction. For ease of presentation,  the firms adjustment costs are assumed to be equal  $$\eta_i = \eta \ \text{ for all } i=1,\ldots,N$$

\subsection{Static allocation}

For the sake of comparison,  consider  an ETS Phase 2-like mechanism,  i.e. a static allocation:  an initial endowment $X^i_0 = \tilde{S}^i_0 =  x^i_0 $ and zero intertemporal allocation  $\tilde a^i_t=0, \tilde S^i_t =0$ for all $i=1,\ldots,N$ and $0< t \leq T$. Under this policy   we now calculate  the social cost \eqref{eq:Reg-OF}, which will necessarily be suboptimal.  Denote by $\bar x_0$ the average initial endowment. Since $a^i_t = - \mu_i$,
by Proposition~\ref{prop:eqmarketnu0}~(i) rewritten with $\eta_i=\eta$ the equilibrium price is
\begin{align} \label{eq:Pets}
\hat P_t & = \frac{2\lambda}{1+2\lambda \eta T } \Big( T \eta \bar h -  \bar x_0  + T \bar \mu  \Big) +  \int_0^t \frac{2 \lambda}{1+2\lambda \eta (T-s)  }      d\bar W_s,
\end{align}
since $\bar M_0 = \bar x_0 - T\bar \mu $. The expected emissions under optimal effort $\hat \a$ given the above price  are 
$$\E\big[  E^{\hat \alpha}_T \big]   =   NT (\bar \mu -   \eta ( \hat P_0 - \bar h) )$$ Then, under a static allocation
the regulator achieves the objective of emission reduction from the BAU trend $T N \bar \mu$ to $\rho T N \bar \mu$ by setting
$$   NT (\bar \mu -   \eta ( \hat P_0 - \bar h) ) = \rho T N \bar \mu. $$
Thus, the initial price $\hat P_0$ becomes identical to \eqref{eq:price-opt-dyn} and the regulator has to allocate  on average
\begin{align}
\bar x_0  = T \rho \bar \mu - \frac{1}{2\lambda} \Big[ \bar h +   \frac{(1-\rho) \bar \mu}{\eta} \Big].
\end{align}

Note that contrary to the optimal dynamic allocation scheme (Theorem~\ref{social-cost-min}~(2)), the initial allocation here can be positive.  Further, let us compare with  the intuitive initial allocation of a cap-and-trade system. There, if regulator wishes to reduce the emissions to $\rho N \bar \mu T$, this value will be the aggregate cap. Then, she would set precisely the initial endowment  at the cap level $\rho  \bar \mu T$ per firm. Here instead the optimal initial endowment  is lower than this intuitive cap.

The corresponding social cost  is:
 \begin{align*}
 C^{\rm stat} & = \E\Big[ \int_0^T \sum_{i=1}^N  \frac12 \eta (   \hat P_t^2  - h_i^2 ) dt + N    \frac{\hat P_T^2}{4\lambda } \Big].
 \end{align*}
 Let denote $d\langle \bar W\rangle_t = \sigma^2 dt$, with $N^2 \sigma^2 :=\sum_{i=1}^N \sigma_i^2 + 2 \sum_{i<j} \rho_{ij} \sigma_i \sigma_j. $ Since
 $$ \E[\hat P^2_t] = \hat P^2_0 +   \int_0^t \left(\frac{2 \lambda}{1+2\lambda \eta (T-s)  } \right)^2 \sigma^2 ds = \hat P_0^2 + \frac{2\lambda \sigma^2}{ \eta}\left( \frac{1}{1+2\lambda \eta (T-t)} - \frac{1}{1+2\lambda \eta T} \right),    $$
straightforward computations lead to
 \begin{align}\label{eq:static-cost}
 {C^{\rm stat} } & =
\frac{N}{4\lambda} \Big( 1+2\lambda\eta T \Big) \hat P^2_0
+     \frac{N \sigma^2}{2\eta} \ln\Big[ 1 +2\lambda\eta T \Big]
  - \frac12 T \eta \sum_{i=1}^N h_i^2
 \end{align}
Thus, the difference in social benefit between the static allocation~\eqref{eq:static-cost} and  the optimal allocation  $\hat C$ in \eqref{eq:social-cost} is  given by
 \begin{align}\label{eq:DeltaStat1}
\Delta^{\text{ {\rm stat}}} & :=
  \frac{N \sigma^2 }{2\eta} \ln\Big[ 1 +2\lambda\eta T \Big]
 \end{align}
Such difference  stems from the presence of uncertainty and inflexibility in the system.  Further, suppose there are $N$ firms with identical $\sigma_i = \bar \sigma$ and identical $k_i$, so that we have $\rho_{ij} = \bar \rho$. \\
\indent In this case,   $N^2 \sigma^2 = N \bar \sigma^2 + \bar \rho \bar \sigma^2 N(N-1)$. Thus, when $N$ is large, the volatility $\sigma^2$ tends to zero, unless there is some correlation with the common shocks.  In fact, when $\bar \rho$ is non zero  the per unit difference in cost does not vanish:
 \begin{align}\label{eq:DeltaStat}
\lim_{N \to \infty}\frac{ \Delta^{\text{ {\rm stat}}} }{N} = \frac{\bar \rho\,  \bar \sigma^2 }{\eta} \ln\Big[ 1 +2\lambda\eta T \Big].
 \end{align}
 Thus, for $N$ large, the optimal dynamic policies continue to  outperform the  static allocation  in  the presence of  common economic shocks. 

Further, by the relation~\eqref{eq:Pets}, the price  quadratic variation satisfies:
\begin{align}\label{eq:volPets}
\langle \hat P \rangle_T = \frac{4 \lambda^2 \sigma^2 T}{1+2\lambda \eta T},
\end{align}
which provides a way to estimate the flexibility parameter $\eta$. Indeed, it satisfies
\begin{align}\label{eq:etaets}
\eta = \frac{  4 \lambda^2  \sigma^2 T-\langle \hat P \rangle_T}{2\lambda T \langle \hat P \rangle_T}.
\end{align}

When the term $\lambda^2 \sigma^2 T$ is large  (see the numerical illustration below in section~\ref{ssec:numeric}), we have
\begin{align}\label{eq:etaets2}
\frac{ \langle P \rangle_T }{\sigma^2} \approx  \frac{2 \lambda}{\eta}.
\end{align}
The relation makes more explicit the relation between the volatility of the exogenous economic shocks and the equilibrium market price volatility. The penalty factor $\lambda$ and the flexibility parameter $\eta$ act as the transmission belts of the economic shocks to the market price volatility. The higher the flexibility $\eta$, the greater the compensation of an economic shock in the equilibrium market price. 

\subsection{Pure tax}

As the name suggests, in a pure tax system there is no bank account for net emissions positions, nor allowances. Firm $i$ makes abatement effort only because of a proportional tax $\tau$ on total realized emissions $E^{i,\alpha^i}_T$ from \eqref{eq:bank-emissions}.  Each firm then faces the minimization problem:

\begin{align}
\inf_{\alpha^i} \E\Big[ \int_0^T c_i(\alpha^i_t)  dt + \tau E_T^{i,\alpha^i}  \Big], \ \quad i=1,\ldots, N
\end{align}
 which  admits a unique solution, the constant effort $\hat \alpha^i = \eta (\tau - h_i)$. So,  the regulator would set the tax at
 \begin{align}
 \tau = \bar h + (1-\rho) \bar \mu/\eta
 \end{align}
   to induce a reduction of expected total emission of a factor $1-\rho$.  Without surprise,  the tax level is equal to the constant price $\hat P_0$ in \eqref{eq:price-opt-dyn} of the dynamic allocation schemes because there the expected emissions reduction is determined by the average carbon price.
The social cost  becomes
\begin{align}
C^{\rm tax} &:=  \E\Big[  \sum_{i=1}^N \int_0^T c_i(\hat \alpha^i_t)  dt + \tau E_T^{i, \hat \alpha^i}  \Big] =    {NT \left ( \frac{\eta}{2} \tau^2   -  \frac{\eta }{2N} \sum_{i=1}^N  h_i^2  +  \rho \bar \mu \tau \right )}
\end{align}
and must be compared with $\hat C$ in \eqref{eq:social-cost}. Quick computations show that the tax is more efficient than an optimal dynamic allocation when
 \begin{align}\label{eq:Ltax}
  \lambda < \frac{\bar h + (1-\rho) \bar \mu/\eta}{4\rho\bar \mu T}.
  \end{align}
The difference in cost between the tax system and the optimal dynamic allocation should be understood as follows. In a tax system, firms pay for each ton they emit whereas in the cap-and-trade system, they pay only for those tons which are non-compliant with the targets. Hence, the more the system emits carbon, the more it is socially expensive to set a tax. The relation~\eqref{eq:Ltax} translates this phenomenon into a threshold on the penalty factor, i.e. the damage function.  

\subsection{MSR--like policy}

To cope with the imbalances of the EUTS between realised emissions and  total allocations, the EU has launched in 2019 a Market Stability Reserve (MSR). The policy rules under MSR specify that when the number of allowances in circulation falls below a certain threshold value (400 million allowances), the regulator auctions off a share of new allowances ($12$\%). Further, if the total number of allowances exceeds another threshold (800 million allowances), the same fraction of allowances are withdrawn from the market.

To make a comparison of the MSR policy with our framework, we consider a continuous time version of this mechanism.  We work under identical initial endowments and allocations, since the social cost is a function of the aggregate quantities. So, consider a net allocation scheme with equal  initial endowment  $X^i_0 = \tilde{S}^i_0 =  x^i_0  = \bar x_0$ (but no intertemporal singular $\tilde S^i_t =0$ for all $i=1,\ldots,N$ and $0< t \leq T$),  and identical  net allocation rates $a^i$ given by:
\begin{align}
a^i_t = \bar a_t =\delta \Big(\frac{T-t}{T} \bar x_0 - \bar X_t \Big), \quad \text{for all } i=1,\ldots,N.
\end{align}

The rationality behind the above allocation mechanism is that the regulator would like to drive the  accounts from $ X^i_0 = \bar x_0 $  to an aggregate position $\bar X_T$ so that $ \E[\bar X_T] \approx 0$  by following a linear trajectory. In this scheme, deviations from the average expected trajectory of carbon emissions reduction are considered to be market imbalances and are compensated continuously at a rate proportional to the imbalance. The parameter $\delta$ acts as mean-reversion factor, trying to make the average bank accounts go back to the desired trajectory. 
 \\
 \indent Since the allocation process is fixed, we need to determine $\bar x_0$ such that
\begin{align}
  \bar x_0 + \E\Big[\int_0^T \bar a_t dt\Big] = \ell(\rho) 
\end{align}
to ensure that the expected total emissions are reduced by a factor $\rho$.
Under this MSR-like mechanism, the market equilibrium follows the dynamics:
\begin{align}
 d\bar X_t & = \left (\eta(\hat P_t -\bar h)   + \delta \Big(\frac{T-t}{T} \bar x_0 - \bar X_t \Big)  \right ) dt - d\bar W_t  \quad  \bar X_0 =\bar x_0,\\
 \hat P_t &  =  F(t)\Big[   (1-\delta z(t)) \Big( \frac{T-t}{T} \bar x_0 - \bar X_t \Big) + z(t)\Big(\eta \bar h  - \frac{\bar x_0}{T} \Big)   \Big],\\
z(t) & = \frac{1-e^{-\delta(T-t)}}{\delta}, \quad F(t) := \frac{f(t)}{1-\eta f(t) \big[T-t - z(t)\big] }, \nonumber
\end{align}
 Since $\bar X$ can be found explicitly,  we can calculate the initial allocation that ensures a reduction of emission of level $\rho$: 
\begin{align}
\bar x_0 = \frac{\delta T}{1-e^{-\delta T}} \Big[ \ell(\rho) + \Big(T + \frac{e^{-\delta T} - 1}{\delta}\Big) \eta \big(\hat P_0 - \bar h\big) \Big].
\end{align}
Computations are  detailed in Appendix~\ref{app:MSR}. The complexity of the  expression for the induced cost leads us to resort to numerical illustrations, see the next Section. 

 \subsection{Numerical illustrations}
 \label{ssec:numeric}

We illustrate here the firms behavior  and provide some orders of magnitude of social costs in the various policy schemes above. As a reference situation, we consider an objective of reduction of 20\% of carbon emission ($\rho$ $=$ $0.8$) over a period of $T=10$ years. \\
\indent This setup is quite close to the objective of the European Union in their climate policy adopted in 2008 for the Phase~2 of the EU~ETS,   only we consider a longer period of time, ten years instead of five.  We consider  the six main sectors covered by the EU~ETS (Public Power and Heat, Pulp and Paper, Cement, Lime and Glass, Metals, Oil and Gas, Other), so $N=6$. The average emission growth rate of the EU 27 members included in the EU~ETS  is around $N \bar \mu = 2$~Gton per year. The average standard deviation of emission rate is $\sigma = 0.2$~Gton/year. \\
\indent The volatility has been estimated on data provided by the European Environment Agency EU Emission Trading System data viewer. For sake of simplicity, we assume that sectors share an identical emission volatility $\sigma_i^2 =   \sigma^2/N$. To estimate the correlation matrix among sectors,  we considered the yearly verified emissions from 2008 to 2012. The result is an average correlation to the common shock of $k_i=0.92$  equal for all sectors.    The terminal penalty parameter $\lambda$ is chosen  so to ensure that, in the optimal dynamic scheme, the reduction target is reached with a discrepancy of $0.1$~Gt. This means that $\lambda$ verifies $N \hat P_0 = 2\lambda |\bar X_T|$ with $|\bar X_T| = 0.1$~Gt, namely $\lambda = 7.5~\,10^{-7}$~\euro/ton$^2$.

The estimation of the marginal abatement cost function per firm is the subject of a vivid, current debate amongst economists (see Gillingham and Stock (2018) \cite{Gillingham18} for an introduction on the topic). Our purpose here is to provide illustrative  examples   to fix the ideas on the difference in social costs across policies. Thus, we base  our choice for $h, \eta$ on the estimation performed by Gollier (2020) \cite{Gollier20} which are in turn based on the MIT Emissions Prediction and Policy Analysis of Morris et. al. (2012) \cite{Morris12}. It leads to taking $N \bar h = 25$~\euro/ton and a nominal value of the flexibility parameter $\eta = 6\,10^{8}$ ton$^2$\euro/year. Further, we make a sensitivity analysis of the costs in function of  $\eta$.

\begin{figure}[!thb]
\begin{flushleft}
\begin{tabular}{c c c}
(a) & (b) & (c) \\
\hspace{-5mm}\includegraphics[width=0.33\textwidth]{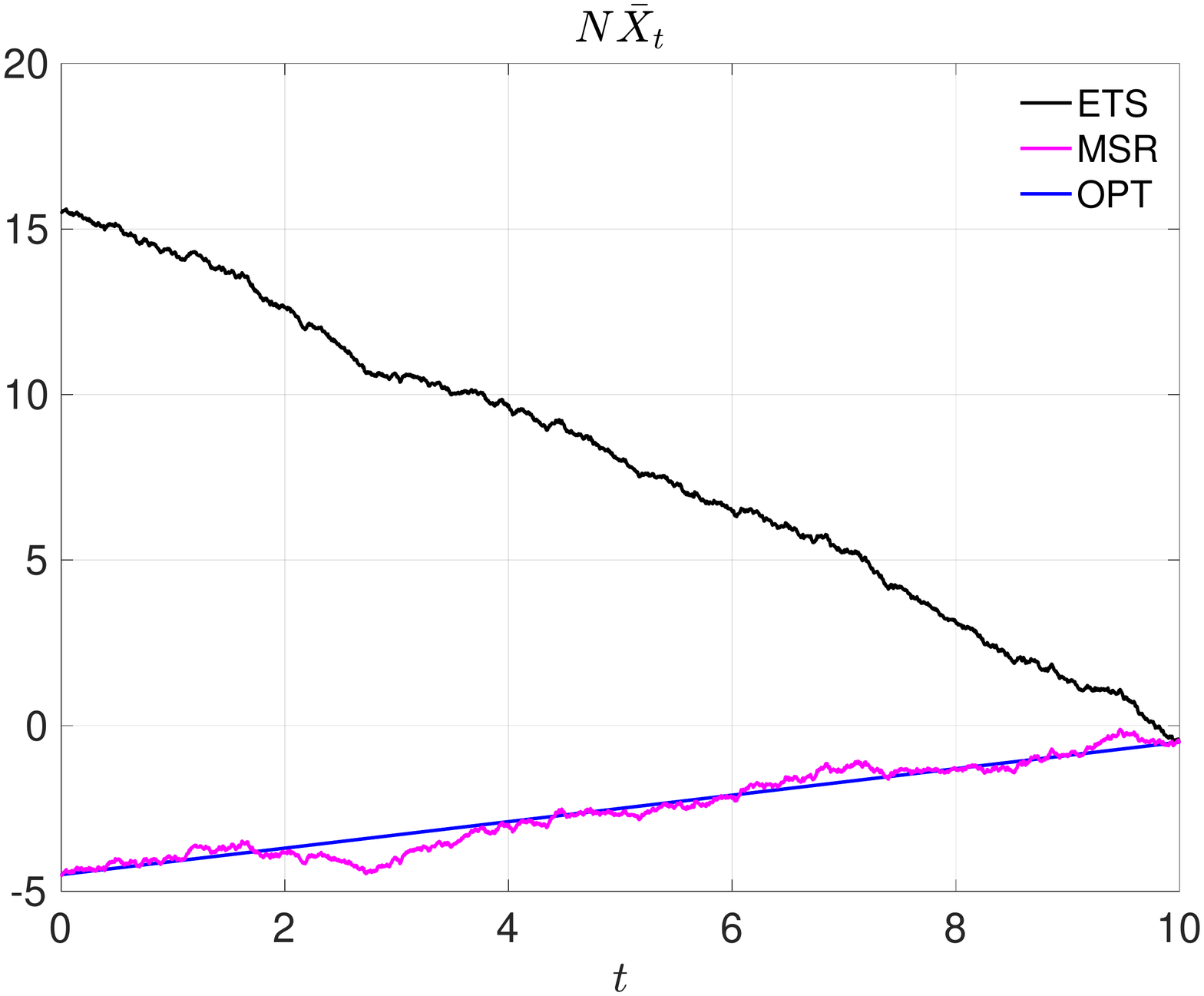}&
\includegraphics[width=0.33\textwidth]{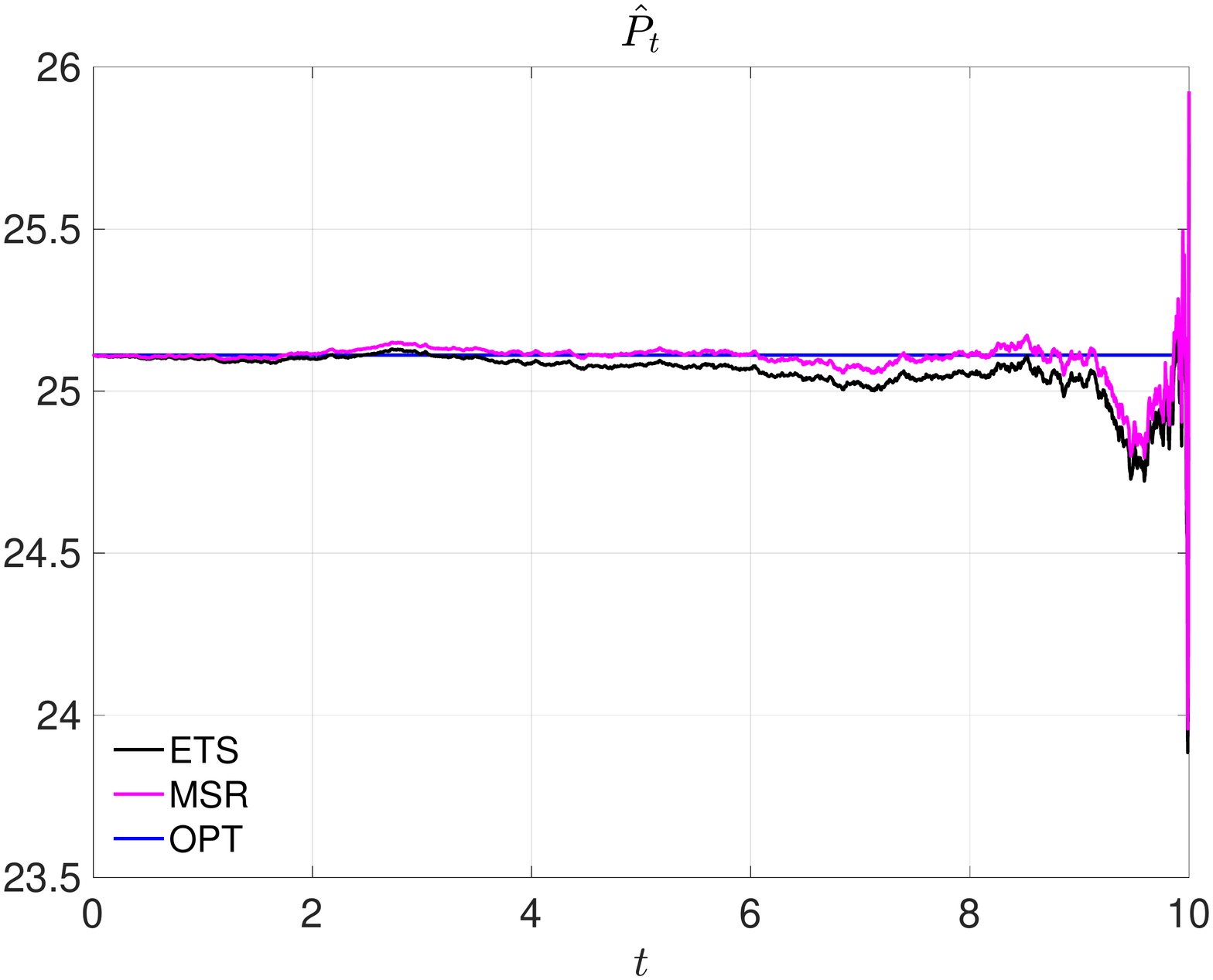}&
\includegraphics[width=0.33\textwidth]{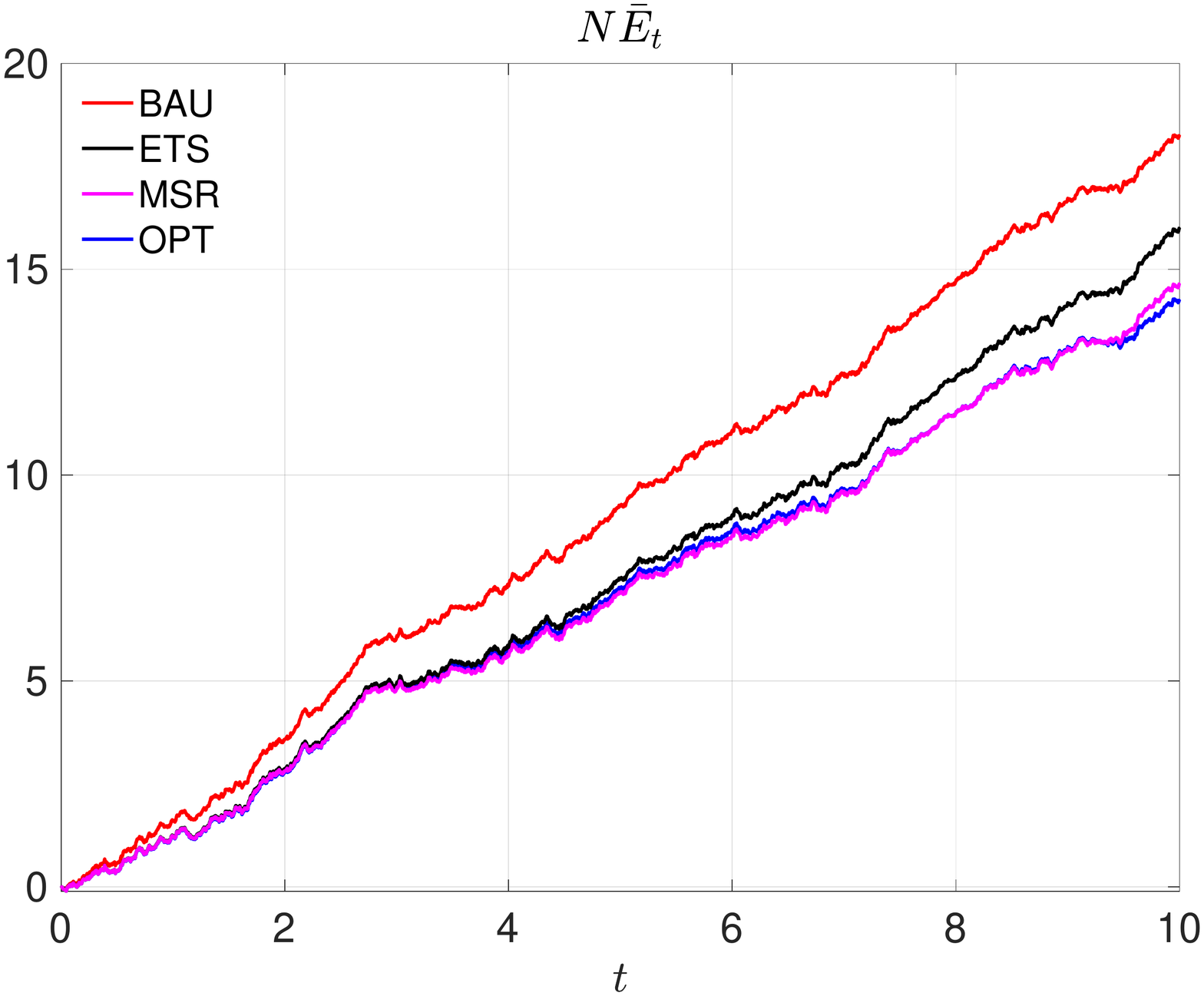}\\
(d) & (e) & (f) \\
\hspace{-5mm}\includegraphics[width=0.33\textwidth]{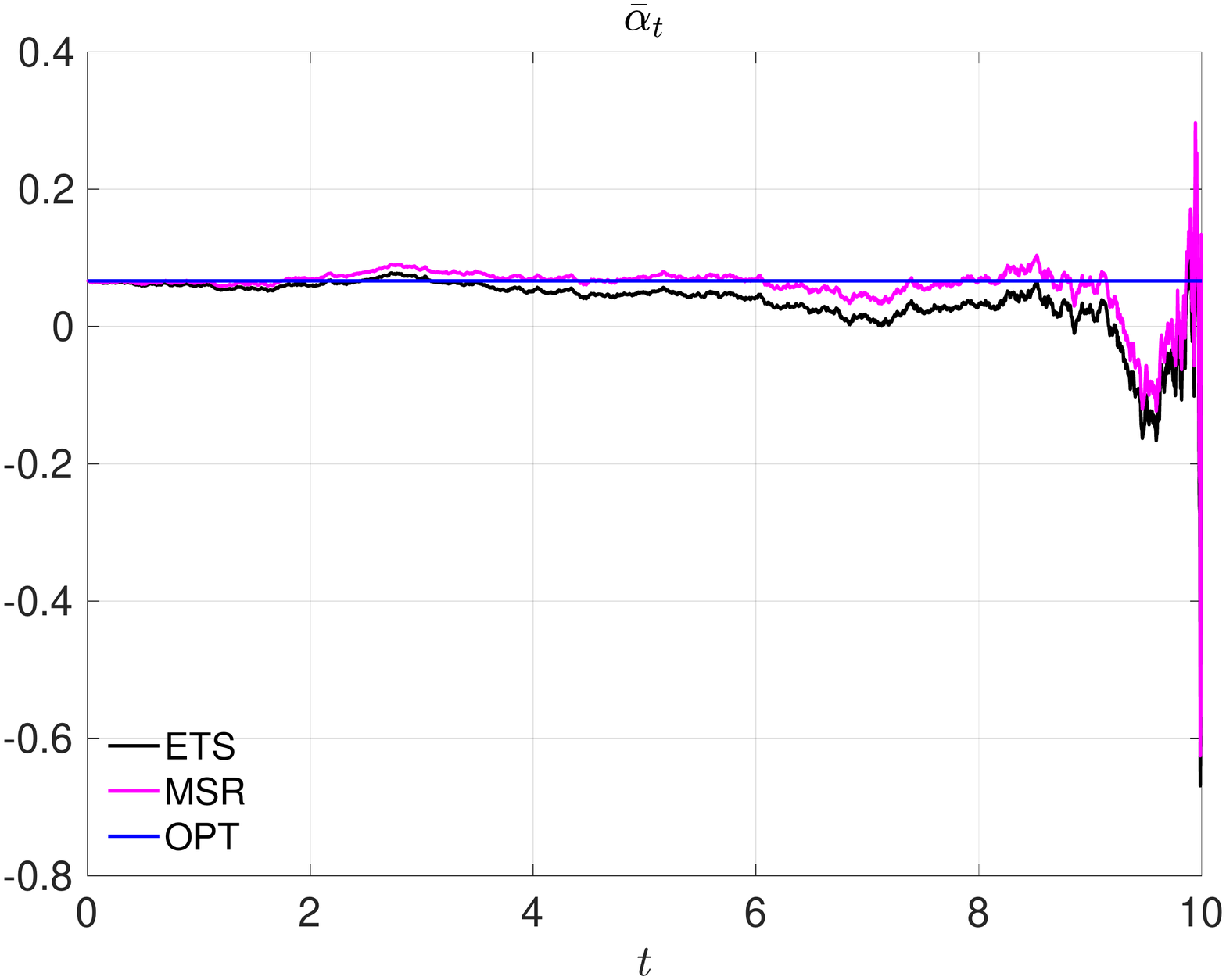}&
\includegraphics[width=0.33\textwidth]{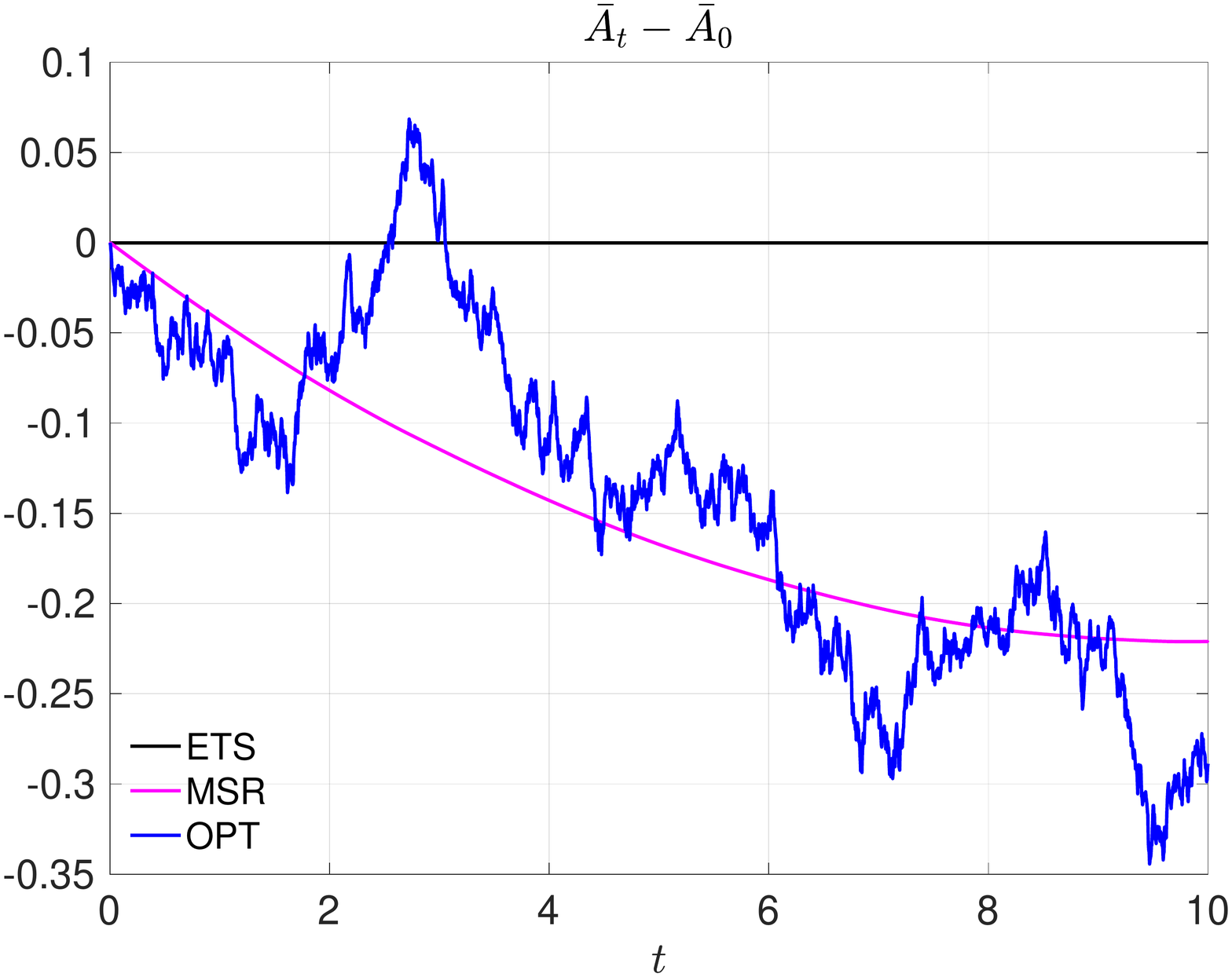}&
\includegraphics[width=0.33\textwidth]{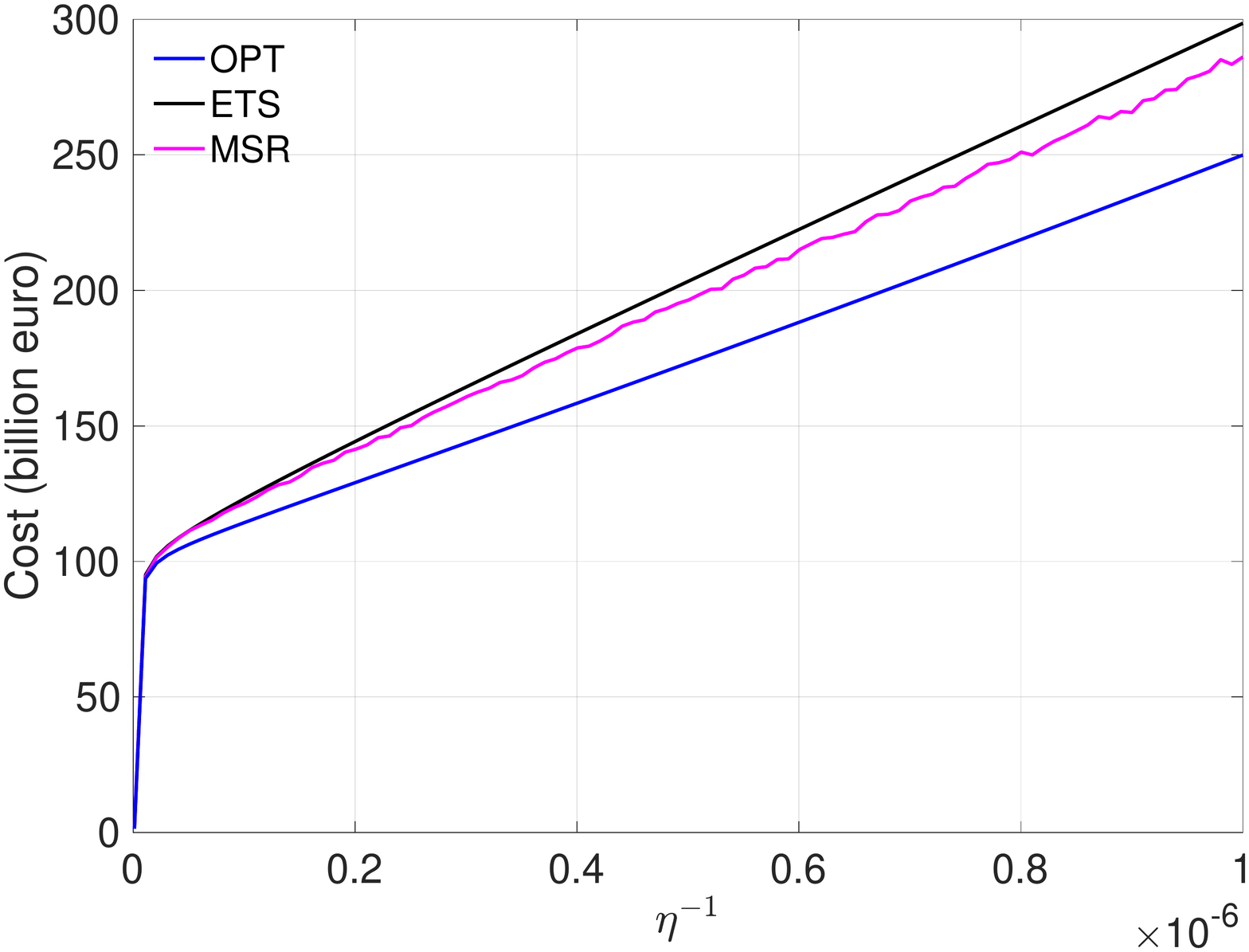}
\end{tabular}
\caption{{\small Simulation of one trajectory under the optimal dynamic policy, the ETS policy and the MSR--like policy of (a) the total bank accounts, (b) the equilibrium market prices, (c) total emission, (d) average abatement effort and  (e) net allocation minus initial allocation. Parameters values: $T=10$ years, $N =6$ sectors, $k_i = k = 0.92$, $\rho = 0.8$, $\mu_i = 2/N$ Gt/year, $\sigma_i = 0.2/\sqrt{N}$~Gt/year, $\bar h = 25$~\euro/t, $\eta=6\,10^8$~ton$^2$/year.\euro, $\lambda = 7.5\,10^{-7}$~\euro/ton$^2$, $\delta =0.1$. Picture (f) Social costs of the different allocation policies as a function of $\eta$; Theorem~\ref{social-cost-min}~(7) for optimal dynamic allocation, formula~\eqref{eq:static-cost} for static (ETS) allocation, Monte-Carlo estimation for the MSR-like mechanism.}}
\label{fig:sim1}
\end{flushleft}
\end{figure}

 Figure~\ref{fig:sim1} illustrates  the behaviour of the different policies in the case of the presence of common shocks. With an average growth rate of 2~Gt per year on a 10 year period with a reduction objective of 20\%, it leads to an expected reduction of 4~Gt. The optimal dynamic allocation schemes starts by allocating a {\em debt} of approximately 5~Gt to the firms while the static ETS scheme allocates approximately 15~Gt in the system, close to the natural cap of $16$~Gt. Both the static and dynamic allocation schemes achieve nearly the same reduction  at the end of the regulated period as the trajectories show. But to reach that end,  each scheme follows  a different path. Under optimal dynamic allocation policies, total bank accounts, total abatement efforts and market equilibrium price are deterministic while the net cumulated allocation is random. Under the static ETS policy the opposite is true, while under the MSR--like policy each  process involved is stochastic. Also, in  the static policy the volatility is monotone increasing  in time. Indeed, the relation~\eqref{eq:Pets} gives that
\begin{align*}
d\langle P \rangle_t = \frac{4\lambda^2 \sigma^2}{\big( 1 + 2 \lambda \eta (T-t) \big)^2} dt.
\end{align*}
Hence, compared to its  value at time zero, the volatility at maturity is multiplied by a $(2\lambda \eta T)^2$ $\approx$ $10^4$, which explains the large oscillations observed at maturity.
Figure~\ref{fig:sim1}~(f) shows the effect the choice $\eta^{-1}$ $\in$ $(10^{-9}, 10^{-6})$  on the different allocation schemes.  First, we observe the hockey stick form of the costs. When $\eta$ is in the range $(10^{8}; 10^{9})$, the system is flexible enough so that there is no significant difference in cost among the various allocation policies. When $\eta$ goes below $10^{8}$, the difference in cost exhibits  a linear growth.   The discrepancy between the static ETS mechanism and the optimal dynamic scheme becomes approximately $20$\% ($50$~billion euros compared to an optimal cost of $250$ billion). The fact that the MSR-like policy succeeds in getting close to the optimal trajectory of the bank accounts translates in a reduced cost compared to the static allocation. Thus, in the range of values for the flexibility of the system we picked, we find significant difference in costs between static and optimal dynamic allocation, but not an order of magnitude. The pure tax scheme is excluded from these comparisons because its cost is one order of magnitude higher than the other policies. In fact, even in  the base scenario of $\eta=6\,10^{8}$ we get an  expected social cost of the allowances policies   around $47$ billion euros, while the tax social cost is greater than $840$ billion euros.

\section{Conclusion}
\label{sec:conclusion}

We find the optimal dynamic allocation processes that achieve a given expected emission reduction of carbon emissions. They  are non-unique, but have the same effect on the system: they induce constant abatement efforts. As a result, the equilibrium market price is also constant. A priori,  the regulator is not pursuing any price control mechanism. A posteriori, however, the desired reduction goals naturally lead to that conclusion. The efficiency of optimal dynamic allocations scheme compared to sub-optimal yet intuitive policies strongly depends on the flexibility of the system and on its dependence on common business cycles. Non-uniqueness in the optimal policy  suggests that within  our framework the regulator can accomplish not only emission reduction at minimal social cost, but also consistently include other features such as the financing of new non-emissive technology. We leave these extensions for future research.

\appendix

\section{Proofs and computations}
The following  facts will be used in the proofs.
\begin{Lemma}\label{Fubini-exp}
Let  $h$ be a process in $\L$ and $Y \in L^2(\Omega, \mathcal{F}_T,P)$. Then
 \begin{align*}
  \E\Big [ Y \int_0^T h_t dt  \Big] = \E \Big[ \int_0^T dt \, h_t \,\E_t[Y]  \Big].
  \end{align*}
 Namely, the scalar product of $Y$ and the pathwise integral $\int_0^T h_tdt$ equals the scalar product in $\L$ of $h$ and the martingale process closed by $Y$, namely $ (E_t[Y])_t$.
\end{Lemma}
\begin{proof}
The proof is one line:
\begin{align*}
  \E\Big[ Y \int_0^T h_t dt \Big] = \E\Big[\int_0^T Y h_t \, dt  \Big] = \int_0^T dt\, \E[ Y h_t ] = \int_0^T dt \,\E[ h_t \,\E_t[Y]] = \E \Big[ \int_0^T dt \, h_t \,\E_t[Y] \Big]
   \end{align*}
 where the equalities from the second onwards follow from Fubini Theorem and from the properties of conditional expectation. An alternative is  to apply integration by parts  to the product of the martingale  $(\E_t[Y])_t$ and  the bounded variation process $\int_0^\cdot h_t dt$.\hfill$\Box$
\end{proof}
\vspace{5mm}
 The proof of the next Lemma is straightforward.
\begin{Lemma}\label{CE-of-Mart-dyn}
Let  $\a \in\L$ be a  martingale.  Then, consider the martingale $M$ closed by $\int_0^T \a_s ds$
$$ M_t: = \E_t \Big[\int_0^T \a_s ds \Big] $$
which also belongs to $\L$.
Then,
$$ M_t = \int_0^t \a_s ds + (T-t)\a_t $$
 and  the dynamics of $M$ are given by
$$ dM_t = (T-t) d \a_t $$
\end{Lemma}
\subsection{Proof of Theorem~\ref{Theo:firm}}
 \label{app:firm}

For ease of notation, we drop the dependence on $i$ of the coefficients and of the controls. We split the cost function  $J$ in two parts,  running cost $C$ and terminal cost $F$:
  \begin{align*}
   J(\alpha, \beta)=\E\Big[  \int_0^T   \Big( h \alpha_t + \frac{\alpha^2_t}{2\eta} + P_t \beta_t + \frac{1}{2 \nu} \beta_t^2 \Big) dt\Big ]+ \E[\lambda X^2_T] = C(\alpha,\beta) + F(\alpha,\beta).
   \end{align*}
 From the structure it is apparent that both $C$ and $F$ are differentiable.
The differential of the running cost
$$C(\alpha, \beta):= \E\Big[  \int_0^T   \Big( h \alpha_t + \frac{\alpha^2_t}{2\eta} + P_t \beta_t + \frac{1}{2 \nu} \beta_t^2 \Big) dt\Big ]$$
is given by  differentiation inside the integral
\begin{align*}
D_\alpha C_t =  h + \frac{\alpha_t}{\eta}, \quad D_\beta C_t = P_t + \frac{\beta_t }{\nu}.
\end{align*}
 In the terminal penalty  $F(\alpha, \beta): = \E[\lambda X^2_T]$,  the bank $X_T$ equals
\begin{align*}
 X_T & = A_T + \int_0^T (\a_t +\b_t)dt - \sigma W_T
 \end{align*}

 If we apply the chain rule inside the expectation in $F$, namely to  $G(\alpha, \beta): = \lambda X^2_T(\alpha,\beta)$ we get the (non adapted) Frechet gradient:
 $$ DG_t = 2\lambda X_T D X_T $$
 in which $DX_T $ is the gradient of $X_T$ w.r.t. $(\alpha, \beta)$.  As $X_T$ is a linear map, $DX_T$ is simply  the constant bidimensional  process
 $$ D X_T = (   1  \quad     1  ).$$

 Given the regularity of the problem, we can differentiate $F =  \E[G(\alpha, \beta)]$ under the expectation side. This means that the Frechet derivative $DF \in \L\times \L$  must verify

\begin{align*}
 \E \Big[\int_0^T DG_t \Big( \begin{array}{c}
                    Y^1_t \\
                    Y^2_t
                  \end{array}\Big) dt\Big] = \E \Big [\int_0^T DF_t \Big( \begin{array}{c}
                    Y^1_t \\
                    Y^2_t
                  \end{array} \Big)  dt \Big]   \quad \forall \, Y^1, Y^2 \in \L
\end{align*}
To find the two components of $DF$,  make the scalar product of e.g. the first derivative $D_\alpha G$ with a generic~$Y^1$:
 \begin{align*}
 \E \Big[\int_0^T D_\alpha G_t Y^1_t dt \Big] = 2\lambda \E \Big[\int_0^T X_T Y^1_t dt \Big] = 2 \lambda \E \Big[\int_0^T \E_t \big[X_T\big] Y^1_t dt \Big],
\end{align*}
 from which we deduce $D_\alpha F_t = 2\lambda \E_t \big[X_T\big] $. Similarly for $D_\beta G$. Finally,
\begin{equation}\label{DF}
DF_t =  \left ( \begin{array}{cc}
             2\lambda \E_t \big[X_T\big] &
            2\lambda \E_t \big[X_T\big]
            \end{array} \right )
\end{equation}
Putting things together, the  differential of the cost function $J$
\begin{align} \label{differentialJ}
  D_\alpha J = h + \frac{\a_t}{\eta} + 2\lambda \E_t\big[X_T\big], \quad
    D_\beta J   =  P_t +\frac{\b_t}{\nu} + 2\lambda \E_t\big[X_T\big]
\end{align}
The FOC equations for the generic agent $i$ write
 \begin{align}\label{eq:foc}
 D_\alpha J =0,  \quad D_\beta J=0.
  \end{align}
 Subtracting the two equations above,
$$ \b_t = \nu(  h + \frac{\a_t}{\eta} - P_t ), $$
which can be substituted into the first equation:
\begin{equation}
\label{alphaeq}
 h + \frac{\a_t}{\eta} +    2 \lambda \E_t \Big[ A_T + \int_0^T ( \a_t + \nu (h+ \frac{\a_t}{\eta} - P_t) )dt \Big]  - 2\lambda \sigma W_t  =0,\  \forall t\geq 0.
\end{equation}
Thus, it is straightforward   that the optimal abatement (if it exists) $\hat \a$ is a martingale, solution of the equation:
 \begin{align} \label{alfaopt}
 \hat \a_t = -\eta\left( h + 2\lambda M_t + 2 \lambda \E_t \Big[ \int_0^T  (( 1+ \frac{\nu}{\eta}) \hat \a_t + \nu(h-P_t) )dt \Big]  - 2\lambda \sigma W_t \right)
 \end{align}

\noindent In particular, using Lemma \ref{CE-of-Mart-dyn} and the definition of $g, M$ in \eqref{aAg}, the initial value of $\hat \a$ is:

\begin{align}\label{alpha0}
\hat\a_0 & =  - g(0) \Big( \frac{1}{2\lambda}  h +    M_0  + \nu \E \Big[ \int_0^T (h - P_t ) dt \Big] \Big).
\end{align}

\noindent To solve for $\hat\a$, we rewrite \eqref{alfaopt} in differential version:
$$ d \hat{\a}_t =   -\eta \Big\{  2 \lambda  d M_t + 2\lambda \Big( 1+ \frac{\nu}{\eta}\Big) d  \E_t \Big[\int_0^T \hat \a_s ds\Big]  + 2 \lambda \nu d \E_t \Big[ \int_0^T (h-P_s) ds\Big]   - 2\lambda \sigma dW_t \Big\}   $$

By Lemma \ref{CE-of-Mart-dyn}, $ d  \E_t \Big[\int_0^T \hat \a_s ds\Big] = (T-t) d \hat{\a}_t $ and thus

$$ (1 +2\lambda(\eta +\nu)(T-t)) d \hat{\a}_t =   -2 \eta \lambda \left(    d M_t  + \nu d \E_t \Big[ \int_0^T (h-P_s) ds \Big]   -   \sigma dW_t \right) $$
or
\begin{align}\label{dyn-hat-alpha}
d \hat{\a}_t  = - g(t) \Big(    d M_t  + \nu d \E_t \Big[ \int_0^T (h-P_s) ds\Big]   -   \sigma dW_t \Big),
 \end{align}
where the term between parentheses is (the differential of) a square integrable  martingale. The Cauchy problem given by the SDE and the initial condition (A.6) uniquely identifies $\hat \a$, since the dynamics of $P$ and $a$ are exogenous here.  The optimal trade $\hat\b_t$ is then
 \begin{align}\label{optrade}
 \hat \b_t = \nu \Big(h + \frac{\hat \a_t}{\eta} - P_t \Big).
\end{align}
   Note that the optimal trade is a martingale of and only if $P$ is a martingale. To conclude, we rewrite the optimal couple as a function of the state $\hat X$. Start again from the first FOC,
 $$ h + \frac{\hat\a_t}{\eta} + 2\lambda \E_t \big[\hat X_T\big]=0  $$
 and rewrite
 $$  h + \frac{\hat \a_t}{\eta} + 2 \lambda \hat X_t + 2 \lambda \E_t \Big[ A_T- A_t + \int_t^T \Big(a_s +  \hat \a_s +\nu \big(h+\frac{\hat \a_s}{\eta} - P_s \big) \Big)ds \Big] =0 $$

 An application of Lemma \ref{CE-of-Mart-dyn} leads to
 $$  h + \frac{\hat \a_t}{\eta} + 2\lambda \Big(1+ \frac{\nu}{\eta} \Big) (T-t) \hat \a_t  + 2 \lambda \hat X_t + 2\lambda R_t + 2 \lambda \E_t \Big[ \int_t^T   + \nu(h-P_s)   ds \Big] =0 $$
 Finally, $\hat \a $ in feedback form reads as
 $$  \hat \a_t = - g(t) \Big( \frac{ h}{2\lambda} +  \hat X_t + R_t  + \nu \E_t \Big[ \int_t^T (h-P_s)  ds \Big] \Big) $$
 and this concludes the proof, since $M_0=R_0$.\hfill$\Box$

\subsection{Proof of Theorem~\ref{Theo:equilibrium}}
 \label{app:equilibrium}

 \noindent {\rm (i)}
 Summing up over all $i$ the relations \eqref{eq:betaopt} and using market clearing condition,
 \begin{align} \label{eq:Palpha}
  N P_t = \sum_{i=1}^N h_i + \frac{\a^i_t}{\eta_i}.
  \end{align}
From Theorem~\ref{Theo:firm}~(i),
 \begin{align} \label{eq:dalpha}
 d\a^i_t  = -  g_i(t)   \Big( dM_t^{i} - \sigma_i dW_t^i - \nu d\E_t\Big[\int_0^T P_s  ds \Big]  \Big).
 \end{align}
 Thus,
\begin{align*}
 N dP_t  = - \sum_{i=1}^N  \frac{g_i(t)}{\eta_i} \Big( dM_t^{i} - \nu  d \E_t\Big[\int_0^T P_s ds\Big] - \sigma_i dW_t^i \Big),
 \end{align*}
 which shows that $P$ must be a martingale. Hence, by Lemma \ref{CE-of-Mart-dyn},
\begin{align*}
 \Big(N -  \sum_i  \frac{g_i(t)\nu(T-t) }{ \eta_i } \Big) d P_t  =  - \sum_{i=1}^N \frac{g_i(t) }{ \eta_i} \Big( dM_t^{i}  -   \sigma_i dW_t^i \Big),
 \end{align*}
which gives the SDE for the equilibrium price $\hat P$ in \eqref{P-EQPRICE}.

For the initial condition of the equilibrium price dynamics, first using the first FOC in \eqref{alphaeq} written for $t=0$ and substituing $\beta^i_0$ as a function of $\alpha^i_0$ and $P_0$, we find that
\begin{align}
\alpha^i_0 & = - g_i(0) \Big( x^i_0 + M^i_0 + h_i \Big( \frac{1}{2\lambda} + \nu T \Big) - \nu T P_0 \Big).
\end{align}
Then substituting the value above in \eqref{eq:Palpha}, we get after tedious computions:
\begin{align}
P_0 &= \frac1N \sum_{i=1}^N \pi_i(0) \Big( \eta_i h_i T -  M^i_0 \Big).
\end{align}

\noindent {\rm (ii)}   The equilibrium price $\hat P$ is given by the market clearing condition
\begin{align*}
\sum_{i=1}^N \hat \beta^i(\hat P(a),a) = 0.
\end{align*}
According to Theorem~\ref{Theo:firm}~(ii), we deduce that:
\begin{align}
\hat P_t =  \frac1N \sum_{i=1}^N (h_i + \frac{\hat \alpha^i_t}{\eta_i}),
\end{align}
from which we deduce that the equilibrium price is a martingale. Recalling that
\begin{align*}
\hat \alpha^i(\hat X^i_t,P) =  - g_i(t) \Big(  \frac{1}{2\lambda}  h_i + \hat X^i_t +   R^i_t + \nu (T-t) h_i - \E_t \Big[ \int_0^T P_s ds \Big] \Big),
\end{align*}
and using the fact that $\hat P$ is a martingale, we get that the equilibrium price is given by:
\begin{align}
\Big[1 -  \frac1N \sum_{i=1}^N \frac{ \nu  g_i(i) (T-t)  }{\eta_i}  \Big] \hat P_t =  \frac1N \sum_{i=1}^N   h_i -\frac{g_i(t)}{\eta_i} \Big(  \frac{1}{2\lambda}  h_i + \hat X^i_t +   R^i_t + \nu (T-t) h_i  \Big).
\end{align}
Hence,
\begin{align}
\Big[1 -  \frac1N \sum_{i=1}^N \frac{ \nu  g_i(i) (T-t)  }{\eta_i}  \Big] \hat P_t
=
 \frac1N \sum_{i=1}^N   h_i \big\{ 1 - \frac{g_i(t)}{\eta_i}  \big( \frac{1}{2\lambda} +  \nu (T-t) \big) \big\} - \frac{g_i(t)}{\eta_i} \Big(   \hat X^i_t +   R^i_t  \Big).
\end{align}

Further,
\begin{align}
 1 - \frac{g_i(t)}{\eta_i}  \big( \frac{1}{2\lambda} +  \nu (T-t) \big) \big\}  & =   g_i(t) (T-t),
 \end{align}
from which
\begin{align*}
\Big[1 -  \frac1N \sum_{i=1}^N \frac{ \nu  g_i(i) (T-t)  }{\eta_i}  \Big] \hat P_t =  \frac1N \sum_{i=1}^N \frac{g_i(t)}{\eta_i} \Big(  \eta_i h_i  (T-t)  -  \hat X_t^i -   R^i_t \Big).
\end{align*}

Hence,
\begin{align*}
 \hat P_t =  \frac1N \sum_{i=1}^N \pi_i(t) \Big(  \eta_i h_i (T-t)   - \hat X_t^i -   R^i_t   \Big).
\end{align*}

\noindent {\rm (iii)} Using relation \eqref{eq:dalpha} and the expression of the dynamics of the equilibrium price, the relations are immediate.
\noindent {\rm (iv)}  Direct consequence  of the initial condition on the controls and on  the price. \hfill$\Box$

\subsection{Computations for the MSR-like mechanism}
\label{app:MSR}

In our MSR-like mechanism,   allocations consists of initial endowment plus net allocation rate $a^i$ of mean-reverting type:
\begin{align}
X^i_0 := \bar x_0, \quad a^i_t = \bar a_t =\delta \Big(\frac{T-t}{T} \bar x_0 - \bar X_t \Big), \quad \text{for all } i=1,\ldots,N.
\end{align}
The goal is finding  $\bar x_0$ to ensure an expected emissions reduction to a factor $\rho$. From Section~\ref{ssec:mainres}, we know that 
\begin{align}\label{eq:app:x0}
\bar x_0 + \int_0^T \E[ \bar a_t ] dt = \ell(\rho),
\end{align}
to ensure the desired reduction. The dynamics of the  average bank account  verifies
\begin{align}\label{eq:app:dX}
 d\bar X_t & = \left (\eta(\hat P_t -\bar h)   + \delta \Big(\frac{T-t}{T} \bar x_0 - \bar X_t \Big)  \right ) dt - d\bar W_t \ \quad \text{ with } \bar X_0 =\bar x_0,
 \end{align}
where we used the expression of the abatement effort rates  $\hat \a^i$ given by \eqref{eq:alpha-no-nu} and the market clearing condition. The   solution is 
 \begin{align} \label{eq:app:Xbar}
  \bar X_t =  {e^{-\delta t}} \bar x_0 + e^{-\delta t}\int_0^t  e^{\delta s} \Big[ \delta \frac{T-s}{T} \bar x_0 + \eta ( \hat P_s- \bar h)\Big] ds -  e^{-\delta t}\int_0^t e^{\delta s}d\bar W_s.
  \end{align}
 Exploiting martingality of $\hat P$,  
\begin{align*}
\E[\bar X_t ] = \Big[ \eta(\hat P_0 - \bar h)  + \frac{\bar x_0}{ T}   \Big] \frac{    1 - e^{-\delta t}}{\delta} + \frac{T-t}{T} \bar x_0.
 \end{align*}

Since the $\hat P_0$ ensuring a reduction to a factor $\rho$ must satisfy $\hat P_0 = \bar h + (1-\rho) \bar \mu/\eta$, the relation~\eqref{eq:app:x0} becomes
 \begin{align*}
   \bar x_0   -     \Big( \eta(\hat P_0 - \bar h)  + \frac{\bar x_0}{ T} \Big)  \Big( T + \frac{e^{-\delta T} - 1}{\delta} \Big)     = \ell(\rho).
\end{align*}
So,
 \begin{align}\label{eq:app:xbar0}
\bar x_0 = \frac{\delta T}{1-e^{-\delta T}} \Big[ \ell(\rho) + \Big(T + \frac{e^{-\delta T} - 1}{\delta}\Big) \eta \big(\hat P_0 - \bar h\big) \Big].
\end{align}

We now solve for the equilibrium price $\hat P $  corresponding to the MSR-like allocation. According to Proposition~\ref{prop:eqmarketnu0}~(1), 
\begin{align*}
\hat P_t = f(t) \Big( (T-t) \eta \bar h - \bar X_t - \bar R_t \Big).
\end{align*}
The residual expected allocation is 
\begin{align*}
\bar R_t = \E_t\Big[ \int_t^T \bar a_s ds \Big]=    \int_t^T \delta \Big( \frac{T-s}{T}\bar x_0 -  \E_t \big[\bar X_s \big] \Big) ds.
\end{align*}
Using the solution of the dynamics of $\bar X$~\eqref{eq:app:Xbar}, we have that
\begin{align*}
 \delta \Big( \E_t\big[ \bar X_s \big]  -   \frac{T-s}{T} \bar x_0 \Big) &  =
 \delta e^{-\delta (t-s) } \Big(  \bar X_t  - \frac{T-t}{T} \bar x_0  \Big)+   \Big[  \eta ( \hat P_t- \bar h) + \frac{\bar x_0}{T} \Big]  \Big( 1 - e^{-\delta (s- t)} \Big),
 \end{align*}

Thus, we have:
\begin{align*}
\bar R_t =  \Big(  \frac{T-t}{T} \bar x_0  - \bar X_t\Big) \Big[ 1 - e^{-\delta(T-t)} \Big]
- \Big( \eta(\hat P_t - \bar h)  + \frac{\bar x_0}{ T}   \Big) \Big[ T - t - \frac{1-e^{-\delta(T-t)}}{\delta} \Big],
\end{align*}
which we rewrite as
\begin{align*}
 \bar R_t =   z(t) \bar a_t +   \Big( \eta \bar h - \frac{\bar x_0}{T} \Big) \big[ T-t - z(t) \big]
-\eta \big[T-t - z(t)\big]  \hat P_t, \quad \text{ with }  z(t) := \frac{1 - e^{-\delta(T-t)} }{\delta}.
\end{align*}
Hence, solving for $\hat P_t$ gives
\begin{align*}
 \hat P_t   & =  F(t) \Big[   (1-\delta z(t)) \Big( \frac{T-t}{T} \bar x_0 - \bar X_t \Big) + z(t)\Big(\eta \bar h  - \frac{\bar x_0}{T} \Big)   \Big], \quad F(t) := \frac{f(t)}{1-\eta f(t) \big[T-t - z(t)\big] }.
\end{align*}
Thus, under this MSR-like mechanism, the market equilibrium follows the dynamics:
\begin{align}
 d\bar X_t & = \left (\eta(\hat P_t -\bar h)   + \delta \Big(\frac{T-t}{T} \bar x_0 - \bar X_t \Big)  \right ) dt - d\bar W_t  \quad  \bar X_0 =\bar x_0,\\
 \hat P_t &  =  F(t)\Big[   (1-\delta z(t)) \Big( \frac{T-t}{T} \bar x_0 - \bar X_t \Big) + z(t)\Big(\eta \bar h  - \frac{\bar x_0}{T} \Big)   \Big],
\end{align}
with $\bar x_0$ given by~\eqref{eq:app:xbar0}.

\end{document}